\documentstyle[12pt]{article}
\advance\voffset by -0.8in
\advance\hoffset by -0.55in
\newcommand{\Bbb}{\bf}
\newcommand{\hhalf}{{\textstyle{1\over 2}}}
\newcommand{\hb}[1]{\hbox to 4truecm{#1\hfil}}
\newcommand{\hbt}[1]{\hbox to 10truecm{#1\hfil}}
\newcommand\ze{{\Bbb Z}}
\newcommand\ce{{\Bbb C}}
\newcommand\re{{\Bbb R}}
\newcommand\Mon{{\Bbb M}}
\newcommand{\one}{{\underline 1}}
\newcommand\Hil{{\cal H}}
\newcommand\V{{\cal V}}
\newcommand\C{{\cal C}}
\newcommand\Cbar{{\bar C}}
\newcommand\Epsilon{{\cal E}}
\newcommand{\Aut}{{\rm Aut}}
\newcommand{\reg}[1]{(\ref{#1})}
\newcommand{\half}{{1\over 2}}
\newcommand{\third}{{1\over 3}}
\newcommand{\Wbar}{{\overline W}}
\begin{document}
\title{Third and Higher Order NFPA Twisted Constructions of Conformal
Field Theories from Lattices}
\author{P.S. Montague \\ Department of Applied Mathematics and
Theoretical Physics\\ University of Cambridge\\
Silver Street\\ Cambridge CB3 9EW\\U.K.}
\maketitle
\begin{abstract}
We investigate orbifold constructions of conformal field theories from
lattices by no-fixed-point automorphisms (NFPA's) $\ze_p$ for $p$
prime, $p>2$, concentrating on the case $p=3$. Explicit expressions
are given for most of the relevant vertex operators, and we consider
the locality relations necessary for these to define a consistent
conformal field theory. A relation to constructions of lattices from
codes, analogous to that found in earlier work in the $p=2$ case which
led to a generalisation of the triality structure of the Monster
module, is also demonstrated.
\end{abstract}
\section{Introduction}
Over the last ten years, conformal field theory has exhibited a
massive development, with diverse applications in string theory and
two-dimensional critical phenomena as well as more abstract areas of
mathematics \cite{Borchvalg,BPZ,Cardy,FLMbook,FLMacadsci,FLMproc}. We
consider, in this paper, an explicit construction of a class of
theories with possible applications in many of these different arenas.

In earlier work \cite{DGMtwisted}, an explicit general construction
was given for $\ze_2$-twisted orbifolds of the FKS lattice theories,
motivated by the fact that one such twisted theory (that corresponding
to the Leech lattice) was shown by
Frenkel, Lepowsky and Meurman (FLM) \cite{FLMbook} to form a natural
module $V^\natural$ for the Monster group $\Mon$. (Note that there are
direct
string-theoretic interpretations in terms of propagation of bosonic
strings on tori and orbifolds thereof.) This construction extended the
results of
previous work \cite{Hollthesis,CorrHoll1,CorrHoll2} by
addressing a normal ordering ambiguity in the twisted sectors,
and, more importantly, full consistency of the
conformal field theory was demonstrated by verification of all the
relevant locality relations.

FLM had shown that the specific features of $V^\natural$ led to the
existence of a ``triality'' involution which, together with the
obvious automorphisms acting on the theory obtained by lifting
automorphisms of the lattice, generated $\Mon$. It was shown
\cite{DGMtrialsumm,DGMtriality} that this triality could be extended
to a larger class of $\ze_2$-twisted theories, by relating the
straight (FKS) and twisted constructions to analogous constructions of
lattices from binary codes, showing that the code structure which had
hitherto been thought of as merely providing a (useful) analogy to
conformal field theory in fact plays a much more fundamental role in
the structure of such theories.
In establishing symmetries of conformal field theories and the
equivalence of potentially distinct theories, our approach helps
provide a tool for a fuller understanding of their structure.

In this paper, we consider $\ze_3$-twisted (no-fixed-point) orbifolds
of the FKS theories specifically, though $\ze_p$-twists for other
primes $p$ are similar, and are given brief consideration. New
features arise that do not occur in the $\ze_2$ case -- in particular we
lose the advantage of having a completely explicit construction.
Some of the expressions derived have been found previously in
string-theory related work \cite{Gato:tvos}, but our results are more
complete, including, for example, discussion of the twisted sector
cocycles.
The
locality relations are considered, and then we go on to describe a
pair of constructions of Eisenstein lattices from ternary codes which
are found to interconnect with the straight and $\ze_3$-twisted
conformal field theory constructions from lattices in a similar way to
binary codes in the $\ze_2$ case.
This promises to allow a similar generalisation of an
analogue of triality to a wider class of theories, since Dong and
Mason have
demonstrated that the $\ze_3$-twisted Leech lattice theory is again
the Monster module $V^\natural$ \cite{ZpOrbifold}, as well as at a
more basic level providing tools for the explicit construction of the
complete class of self-dual theories at central charge 24
\cite{Schell:Venkov,SchellComplete,Schell:seventy}.

Also, the detailed analysis given in this paper for such
relatively simple examples of twisted conformal field theories will be
of use in the explicit calculation of conjectured properties
of such models.

The work of the paper can be seen again as extending that of Corrigan
and Holowood, who discussed the algebra of vertex operators in twisted
conformal field theories under certain assumptions about operator
product expansions. We try to take this further and present a
systematic approach to provide a more complete definition and remove
any need for assumptions by fully exploiting the powerful consequences
of locality -- the verification of which then forms the bulk of the
remaining work.

Much of the physical motivation present for the $\ze_2$ case in
\cite{DGMtwisted} remains true here.
Many of the interesting applications of conformal field theory
involve twisted fields \cite{stringorb1,stringorb2,DFMS},
perhaps the simplest example
being the analysis of critical phenomena in the 2-d Ising model,
where the spin field is a map from an untwisted to a twisted sector.
In superstring models, the set of space-time
fermionic fields is an example of twisted conformal fields, and a complete
analysis of the fermion emission vertex in these theories
would be the analogue of that
provided here for bosonic ``spin'' operators.
Realistic string models also frequently contain complex, {\em e.g.}
$\ze_n$-twisted
world sheet fermionic and bosonic fields, described in terms of
orbifolds or otherwise
\cite{stringorb1,stringorb2,Bluhm1,Bluhm2,Anton1,Anton2}.
A comprehensive description of
these theories, including their explicit construction in terms of
local
conformal fields, and an understanding of the algebras they
form, will undoubtedly be of use in
singling out interesting physical applications.
\section{Conformal Field Theories}
\label{defs}
We begin by defining what shall be meant by a conformal field theory
(see \cite{DGMtriality} for a fuller discussion of the axioms and
their consequences). Essentially, we shall deal with vertex operator
algebras \cite{Borchvalg} or, as physicists would categorise them,
chiral bosonic meromorphic chiral conformal field theories.

A conformal field theory consists of a Hilbert space $\Hil$, a set
$\V=\{V(\psi,z):\psi\in\Hil,z\in\ce\}$ of ``vertex operators'' (linear
maps from $\Hil$ into itself) and a pair of states $|0\rangle$,
$\psi_L\in\Hil$ such that
\begin{itemize}
\item $V(\psi,z)V(\phi,w)=V(\phi,w)V(\psi,z)$, the {\em
locality axiom}. Note that the left hand side is strictly defined only for
$|z|>|w|$, and similarly for the right hand side, and so we are to interpret
this equality in the sense that on taking matrix elements of either side the
resulting meromorphic
functions are analytic continuations of one another. This axiom is
physically reasonable in that we are to interpret the vertex operator
$V(\psi,z)$ as representing the insertion of the state $\psi$ on to the
world sheet of a string at the point $z$. For a bosonic string theory, the
order of such operator insertions must clearly be irrelevant.
\item $V(\psi_L,z)=\sum_nL_nz^{-n-2}$, with
\begin{equation}
[L_m,L_n]=(m-n)L_{m+n}+{c\over{12}}m(m^2-1)\delta_{m,-n}\,,
\end{equation}
for some scalar $c$, known as the {\em central charge}.
\item $V(\psi,z)|0\rangle=e^{zL_{-1}}\psi$, the {\em
creation axiom}.
\end{itemize}
The remaining axioms are technicalities,
listed here only for the sake of completeness.
\begin{itemize}
\item $x^{L_0}$ acts locally with respect to $\V$, {\em i.e.}
$x^{L_0}V(\psi,z) x^{-L_0}$ is local with respect to all the vertex
operators in $\V$. \item The spectrum of $L_0$ is bounded below.
\item $|0\rangle$ is the only state annihilated by $L_0$, $L_{\pm 1}$, {\em
i.e.}
the only su(1,1) invariant state in the theory.
\end{itemize}
If, in addition, the theory is such that
\begin{itemize}
\item $V\left(e^{z^\ast L_1}{z^\ast}^{-2L_0}\psi,1/z^\ast\right)^\dagger$ is
local
with respect to $\V$, {\em i.e.} \cite{DGMtwisted}
\begin{equation}
\label{hermitian}
V\left(e^{z^\ast L_1}{z^\ast}^{-2L_0}\psi,1/z^\ast\right)^\dagger=
V(\overline\psi,z)
\end{equation}
for some antilinear map $\Hil\rightarrow\Hil$,
\end{itemize}
then we say that it is {\em hermitian}.
\section{The Straight Construction in the $Z\!\!\!Z_3$ Context}
In this section, we describe the FKS or straight ({\em i.e.}
untwisted) construction of a conformal field theory from a lattice
which admits a third order NFPA. This is simply a reformulation of the
construction described, for example, in \cite{DGMtwisted,Hollthesis}, but it
serves to introduce our notation.
\subsection{Lattice Structure}
\label{lattice}
Firstly, we will discuss a reformulation of the lattice structure
behind the construction. Anticipating a connection with ternary codes,
we take our lattice to be a $J$-module, for $J$ a ring other than
$\ze$, which admits a natural $\ze_3$ structure. The obvious choice
is the Eisenstein integers $\Epsilon=\{m+n\omega :m,n\in\ze\}$, where
$\omega=e^{2\pi i/3}$, for which multiplication by $\omega$ is a natural
$\ze_3$ symmetry.

The generalisation of
the concept of a
lattice to a ring $J$ is given as follows \cite{ConSlo}.
Let $K$ be the corresponding
field to $J$, {\em e.g.} $K=\re$ for $J=\ze$, $K=\ce$ for
$J=\Epsilon$.
Let $e_1,\ldots e_d\in K^d$ be linearly
independent over $K$. Then the $J$-lattice $\Lambda$ is defined by
\begin{equation}
\Lambda=\{\lambda_1e_1+\ldots +\lambda_de_d:\lambda_i\in J,\,1\leq i\leq d\}
\,.
\end{equation}
Since $J$ is a ring, we see that $\Lambda$ is closed under addition,
subtraction and multiplication by elements of $J$, {\em i.e.} it is a
$J$-module. A choice for $J$ which will be of relevance later
is $J=\ze[\omega_p]$, the ring of cyclotomic integers, where $\omega_p=
e^{2\pi i/p}$, $p$ prime,
{\em i.e.} $\ze[\omega_p]=\{m_0+m_1\omega_p+\ldots
+m_{p-1}{\omega_p}^{p-1}:m_0\,,\ldots,m_{p-1}\in\ze\}$, with
corresponding
field
$K=\ce$. In particular, $\ze[\omega_3]=\Epsilon$.
Because of the natural $\ze_3$ symmetry of $\Epsilon$, an $\Epsilon$-lattice
$\Lambda$ automatically admits a third order automorphism
$\lambda\mapsto\omega\lambda$ (the scalar product on the lattice is taken to be
$(\lambda,\mu)=\lambda\cdot\bar\mu$).

We may define concepts analogous to those defined for $\ze$-lattices.
In particular, the dual lattice $\Lambda^\ast=\{x\in K^d:x\cdot\bar u\in
J\,\,\forall\,\,u\in\Lambda\}$. $\Lambda$ is said to be integral if
$\Lambda\subset\Lambda^\ast$ and self-dual if $\Lambda=\Lambda^\ast$. It is
known that there are no even self-dual $\Epsilon$-lattices. (Note that, for
$\Lambda$ integral, $\lambda\cdot\bar\lambda$ is a real element of $\Epsilon$,
{\em i.e.} $\lambda\cdot\bar\lambda\in\ze$, so that this statement does make
sense.) This result provides a justification for the claim that an
$\Epsilon$-lattice is the appropriate analogue of a ternary code from which to
begin since there
are no self-dual ternary codes whose weights are all multiples of 6
(see {\em e.g.} \cite{PSMcodes}).

We may construct a $2d$-dimensional $\ze$-lattice $\Lambda_R$ from the
$d$-dimensional
$\Epsilon$-lattice $\Lambda$ \cite{ConSlo}.
Suppose $\Lambda$ has generator matrix
$X+iY$, with $X$, $Y$ real ({\em i.e.} $X+iY$ is a $d\times d$ complex matrix
whose rows $\alpha_1,\ldots ,\alpha_d$ comprise a basis for $\Lambda$). Then we
can define the lattice $\Lambda_R$ by specifying its generator matrix to be
\begin{equation}
\left(
\begin{array}{cc}
\sqrt 2X & \sqrt 2Y\\
-{1\over{\sqrt 2}}(X+\sqrt 3Y) & -{1\over{\sqrt 2}}(Y-\sqrt
3X)\end{array}\right)\,,
\end{equation}
with rows $b_1,\ldots ,b_{2d}$.

Define
\begin{eqnarray}
\langle\lambda,\mu\rangle&=&(\lambda,\mu)+(\mu,\lambda)\\
&=&(\lambda,\mu)+\overline{(\lambda,\mu)}\\
&=&2{\rm Re}\,(\lambda,\mu)\,,
\end{eqnarray}
for $\lambda$, $\mu\in\Lambda$. $\langle\cdot,\cdot\rangle$ is a real
symmetric bilinear
form. Clearly $b_i\cdot b_j=2{\rm
Re}\,(\alpha_i\cdot\bar\alpha_j)=\langle
\alpha_i,\alpha_j\rangle=\langle\omega\alpha_i,\omega\alpha_j\rangle$
for $1\leq i$, $j\leq d$ or $d+1\leq i$, $j\leq 2d$,
and for $1\leq i\leq d$, $d+1\leq j\leq 2d$ $b_i\cdot b_j=-{\rm
Re}\,(\alpha_i\cdot\bar\alpha_j)+\sqrt 3{\rm Im}\,(\alpha_i\cdot\bar\alpha_j)=
\langle\alpha_i,\omega\alpha_j\rangle$. Thus, we have a map
$T:\Lambda_R\rightarrow
\Lambda$ given by
\begin{equation}
T\left(\sum_{i=1}^{2d}m_ib_i\right)=\sum_{i=1}^dm_i\alpha_i+
\sum_{i=d+1}^{2d}m_i\omega\alpha_{i-d}\,,
\end{equation}
such that
\begin{equation}
\rho\cdot\gamma=\langle T\rho,T\gamma\rangle\,,
\end{equation}
for $\rho$, $\gamma\in\Lambda_R$. Hence, if we restrict $\Lambda$ to be
integral, then $\rho\cdot\gamma=2{\rm Re}\,(T\rho,T\gamma)\in\ze$ since
$(\cdot,\cdot):\Lambda_R\times\Lambda_R\rightarrow\Epsilon$, {\em
i.e.}
$\Lambda_R$ is integral.
Also $\rho^2=2{\rm Re}\,(T\rho,T\rho)$. But $(T\rho,T\rho)\in
\Epsilon$ and is real, so $(T\rho,T\rho)\in\ze$ and $\rho^2\in 2\ze$.
Hence, $\Lambda_R$ is in fact even if $\Lambda$ is integral.
\subsection{Hilbert Space Structure and Vertex Operators}
We introduce two sets of oscillators $b^i_n$, $\bar b^i_n$,
$1\leq i\leq d$, such that
\begin{eqnarray}
\ [b_m^i,\bar b_n^j]&=&m\delta_{m,-n}\delta^{ij}\nonumber\\
\ [b_m^i,b_n^j]=&0&=[\bar b_m^i,\bar b_n^j]\nonumber\\
{b^i_n}^\dagger&=&\bar b_{-n}^i\\
(p\equiv b^i_0)|\lambda\rangle&=&{\lambda\over\alpha}|\lambda\rangle\,,\qquad
\lambda\in\Lambda\nonumber\\
(\bar p\equiv\bar b^i_0)|\lambda\rangle&=&
{\bar\lambda\over\alpha}|\lambda\rangle\nonumber\\
b^i_n|\lambda\rangle=\bar b^i_n|\lambda\rangle&=&0\,,\qquad n>0\,,\nonumber
\end{eqnarray}
for some integral Eisenstein lattice $\Lambda$ and some
scale factor $\alpha$, fixed later by considerations of self-duality
and locality.

For
\begin{equation}
\psi=\prod_{a=1}^Mb^{i_a}_{-m_a}\prod_{b=1}^N\bar b^{j_b}_{-n_b}
|\lambda\rangle\,,
\label{state}
\end{equation}
we have [see {\em e.g.} \cite{Hollthesis}]
\begin{equation}
V(\psi,z)=:\prod_{a=1}^M{i\over(m_a-1)!}{d^{m_a}\over dz^{m_a}}X^{i_a}_+(z)
\prod_{b=1}^N{i\over(n_b-1)!}{d^{n_b}\over dz^{n_b}}X^{j_b}_-(z)
e^{i{\bar\lambda\over\alpha}\cdot
X_+(z)}e^{i{\lambda\over\alpha}\cdot X_-(z)}:
\sigma_\lambda\,,
\label{vop}
\end{equation}
where
\begin{equation}
X_+(z)=q-ip\ln z+i\sum_{n\neq 0}{b_n\over n}z^{-n}
\end{equation}
and
\begin{equation}
X_-(z)=\bar q-i\bar p\ln z+i\sum_{n\neq 0}{\bar b_n\over n}z^{-n}\,,
\end{equation}
with
\begin{equation}
e^{i{\bar\lambda\over\alpha}\cdot
q+(z)}e^{i{\lambda\over\alpha}\cdot
\bar q}|\mu\rangle=|\lambda+\mu\rangle\,.
\end{equation}
The relation
\begin{equation}
\label{symm}
\hat\sigma_\lambda(\equiv e^{i{\bar\lambda\over\alpha}\cdot
q+(z)}e^{i{\lambda\over\alpha}\cdot
\bar q})\hat\sigma_\mu=(-1)^{\langle\lambda,\mu\rangle/\alpha^2}
\hat\sigma_\mu\hat\sigma_\lambda
\end{equation}
ensures locality \cite{DGMtwisted}.

The third order automorphism $\theta$ in this picture is simply given by
\begin{eqnarray}
\theta b_n^i\theta^{-1}&=&\omega b_n^i\nonumber\\
\theta\bar b^i_n\theta^{-1}&=&\bar\omega\bar b^i_n\\
\theta|\lambda\rangle&=&|\bar\omega\lambda\rangle\nonumber
\end{eqnarray}
(It is easily checked that this is an automorphism, provided that
$\theta\sigma_\lambda\theta^{-1}=\sigma_{\bar\omega\lambda}$. See
\cite{DGMtwisted} for a discussion of the representation of the
cocycle operators.)
The preferred states $|0\rangle$ and $\psi_L$ are given respectively
by the vacuum state $|0\rangle$ and $b_{-1}\cdot\bar
b_{-1}|0\rangle$.
The moments of $V(\psi_L,z)$ are then found to be
\begin{equation}
L_n=:\sum_{m=-\infty}^\infty b_m\cdot\bar b_{n-m}:\,,
\end{equation}
and also to satisfy the Virasoro algebra with $c=2d$. The remaining
axioms are easily checked.
Note that the conformal weight of the state \reg{state} is
given by
\begin{equation}
h_\psi={(\lambda,\lambda)\over{\alpha^2}}+\sum_{a=1}^Mm_a+\sum_{b=1}^Nn_b\,.
\end{equation}
Thus, we see that the theory is meromorphic provided $(\lambda,\lambda)\in
\alpha^2\ze$ for all $\lambda\in\Lambda$. If we impose this condition on
the lattice $\Lambda$, then the theory is just the standard FKS
construction on the even ($\ze$-)lattice $\Lambda_R/\alpha$.

Finally, note that there exists a hermitian structure, with
\begin{equation}
\label{herm}
\overline\psi=(-1)^{L_0}I\psi\,,
\end{equation}
where
\begin{equation}
Ib_nI^{-1}=-\bar b_n\,;\qquad
I\bar b_nI^{-1}=-b_n\,;\qquad I|\lambda\rangle=|-\lambda\rangle\,.
\end{equation}
\section{The $Z\!\!\!Z_3$-Twisted Theory}
As in the $\ze_2$ case described in \cite{DGMtwisted}, the idea
\cite{Hollthesis} is
that we take the projection of the untwisted theory onto the
$\theta=1$ sub-theory, $\Hil_0$, add in two twisted sectors (for a
third order twist), $\Hil_1$ and $\Hil_2$, which form modules for
$\Hil_0$, and then define appropriate vertex operators mapping between
the sectors to make the whole structure a consistent conformal field
theory, as defined in section \ref{defs}.
\subsection{Hilbert Space Structure}
We introduce two sets of twisted oscillators $c^i_r$, $\bar
c^i_r$, $1\leq i\leq d$, $r\in\ze\pm{1\over 3}$. These satisfy
\begin{eqnarray}
\ [c_r^i,c_s^j]&=&[\bar c_r^i,\bar
c_s^j]=r\delta^{ij}\delta_{r,-s}
\nonumber\\
\ [c_r^i,\bar c_s^j]&=&0
\nonumber\\
\ [b_m^i,c_r^j]&=&[b_m^i,\bar c_r^j]=0
\nonumber\\
\ [\bar b_m^i,c_r^j]&=&[\bar b_m^i,\bar c_r^j]=0\,,
\end{eqnarray}
with ${c_r^i}^\dagger=c_{-r}^i$,
${\bar c}_r^{i\dagger}=\bar c_{-r}^i$.

The two twisted sectors $\Hil_1$ and $\Hil_2$ are built up by the
action of the oscillators $c^i_r$, $\bar c^i_r$ respectively on a
representation space for a set of twisted cocycle operators (see
below) annihilated by $c^i_r$ (respectively $\bar c^i_r$) for
$r>0$.
\subsection{Vertex Operators Corresponding to States in the Untwisted
Sector}
Let us concentrate on $\Hil_1$. We will define vertex operators
$V_1(\psi,z):\Hil_1\rightarrow\Hil_1$ corresponding to states $\psi$
in the untwisted sector $\Hil_0$. As in the $\ze_2$ case
\cite{DGMtwisted}, we will define $V_1$ initially by analogy (as in
\cite{Hollthesis}) with the
vertex operators \reg{vop} in $\Hil_0$, which we now relabel $V_0$,
but will then find that we are forced to correct the operators in
order to circumvent a normal ordering problem.

The operators
\begin{equation}
\label{guessVir}
L_n=\sum_{r\in\ze+{1\over 3}}:c_r\cdot c_{n-r}:+{d\over 9}\delta_{n0}
\end{equation}
(with the obvious normal ordering conventions) satisfy the Virasoro
algebra with $c=2d$. We assume, and must check later,
that these are the moments of the
vertex operator $V_1(\psi_L,z)$.

The conformal weight of the state
\begin{equation}
\chi=\prod_{a=1}^Mc^{j_a}_{-r_a}\chi_0\,,
\end{equation}
where $\chi_0$ is an element of the twisted sector ground state, is
then
\begin{equation}
h_\chi={d\over 9}+\sum_{a=1}^Mr_a\,.
\end{equation}
So, for integral conformal weights/a meromorphic representation, we
require both that $d\in 3\ze$ and also that we project the twisted
sector on to states with $\theta=1$, where
\begin{eqnarray}
\theta c_r^i\theta^{-1}=e^{2\pi ir}c^i_r\nonumber\\
\theta\chi_0=e^{-2pi id/9}\chi_0
\end{eqnarray}
will give an extension of the third order automorphism from $\Hil_0$ to
$\Hil_1$.

We define, for the state $\psi$ given by \reg{state},
\begin{equation}
\widetilde V_1(\psi,z)=:\prod_{a=1}^M{i\over{(m_a-1)!}}{d^{m_a}\over{dz^{m_a}}}
C_+^{i_a}(z)
\prod_{b=1}^N{i\over{(n_b-1)!}}{d^{n_b}\over{dz^{n_b}}}
C_-^{j_b}(z)
e^{i{\bar\lambda\over\alpha}\cdot C_+(z)}
e^{i{\lambda\over\alpha}\cdot C_-(z)}:\gamma_\lambda\,,
\end{equation}
by analogy with \reg{vop}, where
\begin{equation}
\label{C}
C_\pm(z)=i\sum_{r\in\ze\pm{1\over 3}}{c_r\over r}z^{-r}
\end{equation}
and the $\gamma_\lambda$ are cocycle operators. (We use the obvious
normal ordering convention.) Note that $\theta$ is defined so that
\begin{equation}
\label{am}
\label{n29}
\theta\widetilde V_1(\psi,z)\theta^{-1}=\widetilde
V_1(\theta\psi,z)\,,
\end{equation}
provided that it has the appropriate action on the twisted cocycles,
{\em i.e.} $\gamma_\lambda=\gamma_{\omega\lambda}$ for all
$\lambda\in\Lambda$.

[Similarly, for $\Hil_2$ we define
\begin{equation}
\widetilde V_2(\psi,z)=:\prod_{a=1}^M{i\over{(m_a-1)!}}{d^{m_a}\over{dz^{m_a}}}
\Cbar_-^{i_a}(z)
\prod_{b=1}^N{i\over{(n_b-1)!}}{d^{n_b}\over{dz^{n_b}}}
\Cbar_+^{j_b}(z)
e^{i{\bar\lambda\over\alpha}\cdot\Cbar_-(z)}
e^{i{\lambda\over\alpha}\cdot\Cbar_+(z)}:\gamma_\lambda\,,
\end{equation}
where $\Cbar_\pm$ is defined following \reg{C}.
If we set
\begin{equation}
\theta\bar c^i_r\theta^{-1}=e^{-2\pi ir}\bar c^i_r
\end{equation}
and
\begin{equation}
\theta\chi_0'=e^{2\pi id/9}\chi_0'\,,
\end{equation}
for $\chi_0'$ in the ground state of $\Hil_2$, then we see that the
analogue of \reg{am} holds. (See \cite{thesis} (page 132) for a
discussion of the necessity of introducing a second twisted sector.]

We now rewrite $\widetilde V_1$ in the form of a matrix element,
allowing us both to see clearly a problem with its commutation
relation with $L_{-1}$ and, at the same time, to correct it. We try to
write
\begin{equation}
\label{melt}
\widetilde V_1(\psi,z)=\sum_{\mu\in\Lambda}\langle\mu|:
e^{B_1(-z)}:|\psi\rangle\gamma_\mu\,,
\end{equation}
(which is linear in $\psi$, as required), where
\begin{equation}
:
e^{B_1(z)}:=e^{B_1^-(z)}e^{B_1^+(z)}
\end{equation}
and
\begin{eqnarray}
B_1^+(z)&=&\sum_{n\geq 0\atop{r>0}}B^+_{nr}(z)c_r\cdot\bar b_n+
\sum_{n\geq 0\atop{s>0}}B^+_{ns}(z)c_s\cdot b_n\nonumber\\
B_1^-(z)&=&\sum_{n\geq 0\atop{r>0}}B^-_{nr}(z)c_{-r}\cdot b_n+
\sum_{n\geq 0\atop{s>0}}B^-_{ns}(z)c_{-s}\cdot\bar b_n\,.
\end{eqnarray}
[From now on, we shall use $r$ to indicate an element of $\ze+{1\over 3}$, $s$
to indicate an element of $\ze-{1\over 3}$ and $t$ for an element of $\ze\pm
{1\over 3}$.] \reg{melt} is explicitly normal ordered, and a trivial
calculation, analogous to that given in \cite{DGMtwisted}, gives
\begin{equation}
B^\pm_{nt}(z)=\mp{1\over t}\left({\mp t\atop n}\right) (-z)^{-n\mp
t}\,.
\end{equation}
[The expression for $\widetilde V_2(\psi,z)$ is obtained by making the
substitution
$c\rightarrow\bar c$ and interchanging $b$ and $\bar b$.]

Now, it is a simple consequence \cite{DGMtwisted} of the axioms laid
down in section \ref{defs} that the vertex operators $V(\psi,z)$ of a
conformal field theory satisfy
\begin{equation}
\label{deriv}
[L_{-1},V(\psi,z)]={d\over dz}V(\psi,z)\,.
\end{equation}
Following the calculation of \cite{DGMtwisted} for the $\ze_2$ case,
we find that this is not true of the $\widetilde V_1(\psi,z)$ (with
the proposed expression for $L_{-1}$ in the twisted sector). The
problem is that
\begin{equation}
[L_{-1},c^i_r]=-rc^i_{r-1}\,,
\end{equation}
and so terms involving $c^i_{{1\over 3}}$ and $c^i_{{2\over 3}}$ then
present problems, in that they produce additional unwanted terms when
normal ordered after commutation on passing through $c^i_{{2\over 3}}$
and $c^i_{{1\over 3}}$ respectively. This is corrected in a completely
analogous way to that in the $\ze_2$ case, by defining
\begin{equation}
\label{corr}
V_1(\psi,z)=\widetilde V_1\left(e^{A_1(-z)}\psi,z\right)\,,
\end{equation}
where
\begin{eqnarray}
A_1(z)&=&\sum_{n,m\geq 0}A^1_{nm}(z)b_n\cdot\bar b_m\nonumber\\
&=&\sum_{n,m\geq 0\atop{n+m>0}}{1\over{n+m}}\left({-{2\over 3}\atop n}
\right)\left({-{1\over 3}\atop m}\right)b_n\cdot\bar b_m(-z)^{-n-m}
-\ln (-Az)p\cdot\bar p\,,
\end{eqnarray}
and $A$ is an arbitrary constant (of integration), to be later fixed by
consideration of the representation property for the twisted sectors, {\em
i.e.} the relevant intertwining relations. (Note that, though this may
be equivalent to the uncorrected form for $V_1$ with a suitable
redefinition of normal ordering, it has the advantage of giving an
explicit form for the vertex operators -- a property which proves to
be invaluable in the suvbsequent calculations.

The expression for $V_2(\psi,z)$ is again found by interchanging $b$
and $\bar b$. (Note that we use the same constant $A$ in both
cases,
as indeed we use the
same set of twisted cocycles. That this is consistent will become
clear.)

A full understanding of why this procedure works, {\em i.e.} why
merely correcting the $L_{-1}$ commutation relation is sufficient to
give the correct form for the vertex operators, remains to be found.

Note that $V_1(\psi_L,z)=\widetilde V_1(\psi_L,z)+{d\over 9}z^{-2}$,
with modes as given in \reg{guessVir}, justifying our earlier
assumption.

The twisted conformal field theory which we are constructing is built
upon the Hilbert space
$\widehat\Hil\equiv\Hil_0\oplus\Hil_1\oplus\Hil_2$ (on which
$\theta=1$). We postulate that the vertex operator corresponding to
the state $(\psi,0,0)$ (often written as $\psi$ for notational
convenience) is
\begin{equation}
\V(\psi,z)=
\left(
\begin{array}{ccc}
V_0(\psi,z) & 0 & 0\\
0 & V_1(\psi,z) & 0\\
0 & 0 & V_2(\psi,z)
\end{array}
\right)\,.
\end{equation}
It is easily demonstrated ({\em c.f.} \cite{DGMtwisted}) that we have
a hermitian structure on these operators (provided the twisted
cocycles behave appropriately under hermitian conjugation), {\em i.e.}
\begin{equation}
\V(\overline\psi,z)=
\V\left(e^{z^\ast L_1}{z^\ast}^{-2L_0}\psi,1/z^\ast\right)^\dagger\,,
\end{equation}
where $\overline\psi$ is given by \reg{herm}.
\subsection{Remaining Vertex Operators}
In order to complete the verification of $\widehat\Hil$ as a conformal
field theory, we must define vertex operators corresponding to a
general state, {\em i.e.} including non-zero components in the twisted
sectors.

We write
\begin{equation}
\V\left((\psi,\chi,\chi'),z\right)=
\left(
\begin{array}{ccc}
V_0(\psi,z) & \Wbar_1(\chi',z) & \Wbar_2(\chi,z)\\
W_1(\chi,z) & V_1(\psi,z) & \Wbar_3(\chi',z)\\
W_2(\chi',z) & W_3(\chi,z) & V_2(\psi,z)
\end{array}
\right)\,.
\end{equation}
If the resulting theory is to have a hermitian structure, we must have
\begin{equation}
\label{n789}
\Wbar_i(\overline\chi,z)=
W_i\left(e^{z^\ast L_1}{z^\ast}^{-2L_0}\chi,1/z^\ast\right)^\dagger\,,
\end{equation}
for $i=1$, 2, 3 and some anti-linear involution
$\chi\mapsto\overline\chi$ from $\Hil_1\rightarrow\Hil_2$.

We will obviously take the preferred states to be $(|0\rangle,0,0)$ and
$(\psi_L,0,0)$. The creation axiom then simply gives us
\begin{equation}
\label{cr1}
V_0(\psi,z)|0\rangle=e^{zL_{-1}}\psi\,,
\end{equation}
which holds as it is satisfied in the untwisted conformal field
theory,
\begin{equation}
\label{cr2}
W_1(\chi,z)|0\rangle=e^{zL_{-1}}\chi
\label{n46}
\end{equation}
and
\begin{equation}
\label{cr3}
W_2(\chi',z)|0\rangle=e^{zL_{-1}}\chi'\,.
\end{equation}
The only remaining axiom to verify, and which must furnish us with
sufficient information to define $W_1$, $W_2$ and $W_3$ (and the
conjugation map on the twisted states), is the locality axiom.
The locality relations for the operators $\V(\rho,z)$,
$\rho\in\widehat\Hil$, are seen to be equivalent to 18 distinct
locality relations for the component vertex operators. Many of these
are hermitian conjugates of others. We consider the 12 that remain,
{\em i.e.}
\begin{eqnarray}
\label{one}
V_0(\psi,z)V_0(\phi,w)&=&V_0(\phi,w)V_0(\psi,z)\\
\label{two}
V_1(\psi,z)V_1(\phi,w)&=&V_1(\phi,w)V_1(\psi,z)\\
\label{three}
V_2(\psi,z)V_2(\phi,w)&=&V_2(\phi,w)V_2(\psi,z)\\
\label{four}
W_1(\chi,z)V_0(\psi,w)&=&V_1(\psi,w)W_1(\chi,z)\\
\label{five}
W_2(\chi',z)V_0(\psi,w)&=&V_2(\psi,w)W_2(\chi',z)\\
\label{six}
\Wbar_1(\chi',z)W_1(\chi,w)&=&\overline
W_2(\chi,w)W_2(\chi',z)\\
\label{seven}
W_3(\chi,z)V_1(\psi,w)&=&V_2(\psi,w)W_3(\chi,z)\\
\label{eight}
W_3(\chi,z)W_1(\rho,w)&=&W_3(\rho,w)W_1(\chi,z)\\
\label{nine}
\Wbar_3(\chi',z)W_3(\chi,w)&=&W_1(\chi,w)\overline
W_1(\chi',z)\\
\label{ten}
W_3(\chi,z)\overline W_3(\chi',w)&=&W_2(\chi',w)\Wbar_2(\chi,z)\\
\label{eleven}
\Wbar_2(\chi,z)W_3(\rho,w)&=&\Wbar_2(\rho,w)W_3(\chi,z)\\
\label{twelve}
W_1(\chi,z)\Wbar_2(\rho,w)&=&W_1(\rho,w)\Wbar_2(\chi,z)
\,,
\end{eqnarray}
for $\psi$, $\phi\in\Hil_0$, $\chi$, $\rho\in\Hil_1$ and $\chi'$,
$\rho'\in\Hil_2$. We know \reg{one} from the fact that the $V_0$'s
specify a sub-theory of a consistent conformal field theory.

Consider the action of \reg{four} and \reg{five} on $|0\rangle$, and
use
(\ref{cr1}-\ref{cr3}). We find (taking $w=0$)
\begin{equation}
W_1(\chi,z)\psi=V_1(\psi,0)e^{zL_{-1}}\chi\,,
\end{equation}
or
\begin{equation}
W_1(\chi,z)\psi=e^{zL_{-1}}V_1(\psi,-z)\chi
\label{W1}
\end{equation}
from (exponentiation of) \reg{deriv}, and similarly
\begin{equation}
W_2(\chi',z)\psi=e^{zL_{-1}}V_2(\psi,-z)\chi'\,.
\end{equation}
Thus, $W_1$ and $W_2$ are fixed uniquely by the requirements of
locality, as was the intertwiner in the $\ze_2$ case. The definition
of $W_3$, the intertwiner of the twisted sectors, is however more
intricate,
and we will discuss this new feature in a later section.

In the following sections, we will consider the above locality
relations in some detail.
\subsection{Representation Property}
Firstly, we will verify (\ref{two}-\ref{five}).

We show that \reg{two} follows from \reg{four}.
Consider
\begin{eqnarray}
V_1(\psi,z)V_1(\phi,w)\chi&=&V_1(\psi,z)V_1(\phi,w)W_1(\chi,0)|0\rangle\nonumber\\
&=&W_1(\chi,0)V_0(\psi,z)V_0(\phi,w)|0\rangle\qquad\hbox{\rm by
\reg{four}}
\nonumber\\
&=&W_1(\chi,0)V_0(\phi,w)V_0(\psi,z)|0\rangle\qquad\hbox{\rm by
\reg{one}}
\nonumber\\
&=&V_1(\phi,w)V_1(\psi,z)W_1(\chi,0)|0\rangle\qquad\hbox{\rm by
\reg{four}}
\nonumber\\
&=&V_1(\phi,w)V_1(\psi,z)\chi\,,
\end{eqnarray}
as required. Similarly, \reg{three} is a consequence of \reg{five}. We
therefore check only \reg{four} and \reg{five}.

[Note that \cite{DGMtriality}
\begin{eqnarray}
V_1(\psi,z)V_1(\phi,w)\chi&=&W_1(\chi,0)V_0(\psi,z)
V_0(\phi,w)|0\rangle\qquad\hbox{\rm from above}\nonumber\\
&=&W_1(\chi,0)V_0\left(V_0(\psi,z-w)\phi,w\right)|0\rangle\nonumber\\
\noalign{\centerline{by the duality property of the conformal field theory
$\Hil_0$ (see {\em e.g.} \cite{PGmer})}}\nonumber\\
&=&V_1\left(V_0(\psi,z-w)\phi,w\right)W_1(\chi,0)|0\rangle
\qquad\hbox{\rm by
\reg{four}}
\nonumber\\
&=&V_1\left(V_0(\psi,z-w)\phi,w\right)\chi\,,
\end{eqnarray}
{\em i.e.}
\begin{equation}
V_1(\psi,z)V_1(\phi,w)=V_1\left(V_0(\psi,z-w)\phi,w\right)\,,
\label{duality}
\end{equation}
and we say that $\Hil_1$ is a representation of the conformal field
theory $\Hil_0$ \cite{DGMtriality}. Similarly, $\Hil_2$ also forms a
module for $\Hil_0$.]

Consider now the locality relation \reg{four}. We write
$V_0(\psi,z)$ in matrix element form:-
\begin{equation}
\label{melt2}
V_0(\psi,z)=\sum_{\mu'\in\Lambda}\langle\mu'|e^{i\bar p'\cdot q}
e^{ip'\cdot\bar q}:e^{M(z)}:|\psi'\rangle\sigma_\mu'\,,
\end{equation}
where
\begin{equation}
:e^{M(z)}:=e^{M^-(z)}e^{M^+(z)}\,,
\end{equation}
with
\begin{equation}
M^+(z)=\sum_{n,m\geq 0}M^+_{nm}(z)b'_n\cdot\bar b_m+
\sum_{n,m\geq 0}M^+_{nm}(z)\bar b'_n\cdot b_m
\end{equation}\
\begin{equation}
M^-(z)=\sum_{{n\geq 0}\atop{m>0}}M^-_{nm}(z)b'_n\cdot\bar b_{-m}+
\sum_{{n\geq 0}\atop{m>0}}M^-_{nm}(z)\bar b'_n\cdot b_{-m}\,,
\end{equation}
the $'$ indicating a second (commuting) copy of $\Hil_0$. A
simple argument ({\em c.f.} \cite{DGMtwisted}) gives
\begin{equation}
M^\pm_{nm}(z)=
\mp{1\over m}\left({\mp m\atop n}\right) z^{\mp m-n}\,,
\end{equation}
for $n\geq 0$, $m>0$, and
\begin{equation}
M^+_{00}(z)=\ln z
\end{equation}
\begin{equation}
M^+_{n0}(z)=-{1\over n}(-z)^{-n}\,,
\end{equation}
for $n>0$.

Suppose that
\begin{equation}
\gamma_\lambda\gamma_\mu=\epsilon(\lambda,\mu)e^{iS(\lambda,\mu)}
\gamma_{\lambda+\mu}\,,
\end{equation}
where
\begin{equation}
\label{n745}
\hat\sigma_\lambda\hat\sigma_\mu=\epsilon(\lambda,\mu)\hat\sigma_{
\lambda+\mu}\,,
\end{equation}
{\em i.e.} we allow for a possible difference in the symmetry factors for the
straight and twisted sector cocycle operators. This gives
\begin{equation}
\gamma_\lambda\gamma_\mu={\epsilon(\lambda,\mu)\over{\epsilon(\mu,
\lambda)}}e^{P(\lambda,\mu)}\gamma_\mu\gamma_\lambda\,,
\end{equation}
where
\begin{equation}
P(\lambda,\mu)\equiv i\left(S(\lambda,\mu)-S(\mu,\lambda)\right)\,.
\end{equation}
Thus,
\begin{equation}
\label{gammaT}
\gamma_\lambda\gamma_\mu=(-1)^{\langle\lambda,\mu\rangle/\alpha^2}
e^{P(\lambda,\mu)}
\gamma_\mu\gamma_\lambda\,,
\end{equation}
using \reg{n745} together with \reg{symm}.

{}From \reg{W1}, \reg{corr} and \reg{melt} we have
\begin{equation}
W_1(\chi,z)=e^{zL_{-1}}\sum_{\mu\in\Lambda}\langle\mu|:e^{B_1(z)}:
e^{A_1(z)}|\chi\rangle\gamma_\mu\,.
\end{equation}
Hence \reg{four} becomes
\begin{equation}
\sum_{\mu\in\Lambda}\gamma_\mu\langle\mu|:e^{B_1(z)}:
e^{A_1(z)}|\chi\rangle V_0(\psi,z)=V_1(\psi,w-z)
\sum_{\mu\in\Lambda}\gamma_\mu\langle\mu|:e^{B_1(z)}:
e^{A_1(z)}|\chi\rangle\,,
\end{equation}
using (exponentiation of) \reg{deriv}, or from \reg{melt2}, \reg{corr}
and \reg{melt}
\begin{eqnarray}
\label{assert}
&&\sum_{\mu\in\Lambda}\gamma_\mu\langle\mu|
:e^{B_1(z)}:e^{A_1(z)}|\chi\rangle\sum_{\mu'\in\Lambda}\langle\mu'|
e^{i\bar p'\cdot q}e^{ip'\cdot\bar q}:e^{M(w)}:
|\psi'\rangle\sigma_{\mu'}\nonumber\\
&=&\sum_{\mu'\in\Lambda}\gamma_{\mu'}\langle\mu'|:e^{B_1(z-w)}:e^{A_1(z-w)}
|\psi'\rangle
\sum_{\mu\in\Lambda}\gamma_\mu\langle\mu|
:e^{B_1(z)}:e^{A_1(z)}|\chi\rangle\,.
\end{eqnarray}
We simply reorder the exponentials on the right-hand side by use of
the Baker-Campbell-Hausdorff identities (see \cite{DGMtwisted} for an
analogous calculation in the $\ze_2$ case).
Then \reg{assert} is found to be equivalent to the following
requirements:-
\begin{equation}
\label{n748a}
[B_1^\pm(\zeta),\widetilde M^-(z)]=B_1^\pm(-(z-\zeta))'
\end{equation}
\begin{equation}
\label{n748b}
\half[\widetilde M^-(-z),[A_1(\zeta),\widetilde M^-(-z)]]=A_1(-(z-\zeta))'
\end{equation}
\begin{equation}
\langle\lambda'|\langle\mu|\left\{
[B_1^-(\zeta),[B_1^+(\zeta),\widetilde M^-(-z)]]
+[A_1(\zeta),\widetilde M^-(-z)]+M^+(-z)\right\}
=\langle\lambda'|\langle\mu|iS(\lambda,\mu)\,,
\label{n748c}
\end{equation}
where
\begin{equation}
\widetilde M^-(z)=M^-(z)+i\bar p'\cdot q+ip'\cdot\bar q
\end{equation}
and
$B_1^\pm(z)'$, $A_1(z)'$ denote $B_1^\pm(z)$, $A_1(z)$ respectively written in
terms of the primed oscillators.

Note the occurrence of the phase factor $S(\lambda,\mu)$ in \reg{n748c}. In the
$\ze_2$-twisted case, we were able to choose the arbitrary constant of
integration in our definition of $A(z)$ such that this term was absent.
However, in this case we see that in general we shall not be able to choose
this constant to cancel both the terms in $p'\cdot\bar p$ and those in $\bar
p'\cdot p$ inside the $\{\ldots\}$ of \reg{n748c}.
What we shall do is to choose
$A$ so that $S(\lambda,\mu)$ is real, which is consistent with a
representation of the twisted cocycles by unitary matrices.

The identities (\ref{n748a}-\ref{n748c}) may be checked by replacing
$b_n$ by $x^n$, $b'_n$ by $y^n$, $\bar b_n$ by $\bar x^n$ and $\bar b'_n$ by
$\bar y^n$ and then summing to obtain the generating functions for the
coefficients of the terms involving these oscillators. [Note that, in
general, this procedure will be unable to distinguish between say
${b_{-1}}^n$ and $b_{-n}$, though because we know the form of the
operators it is sufficient in this particular case.]
The simple identities
which are required are
\begin{eqnarray}
\label{n750a}
\sum_{n,m\geq 0}x^my^nA^1_{nm}(z)&=&-(1-\omega)\ln
\left({(y-z)^\third-\omega (x-z)^\third\over{(1-\omega)}}\right)\nonumber\\
&&\hskip30pt-(1-\bar\omega)\ln
\left({(y-z)^\third-\bar\omega (x-z)^\third\over{(1-\bar\omega)}}\right)
-\ln A
\end{eqnarray}
\begin{equation}
\sum_{n\geq 0}y^nB^\pm_{nt}(z)=\mp{1\over t}(y-z)^{\mp t}
\end{equation}
\begin{equation}
\sum_{r>0}{y^r\over r}=-\ln (1-y^\third)-\bar\omega\ln (1-\omega
y^\third)-\omega\ln (1-\bar\omega
y^\third)
\end{equation}
\begin{equation}
\sum_{s>0}{y^s\over s}=-\ln (1-y^\third)-\omega\ln (1-\omega
y^\third)-\bar\omega\ln (1-\bar\omega
y^\third)\,.
\label{n750d}
\end{equation}
\reg{n748a} and \reg{n748b} are easily checked, while \reg{n748c} gives
\begin{eqnarray}
iS(\lambda,\mu)&=&{\lambda\cdot\bar\mu\over{\alpha^2}}
\{-\ln A+(1-\bar\omega)\ln (1-\omega)
+(1-\omega)\ln (1-\bar\omega)\}\nonumber\\
&&\hskip20pt+{\bar\lambda\cdot\mu\over{\alpha^2}}
\{-\ln A+(1-\omega)\ln (1-\omega)
+(1-\bar\omega)\ln (1-\bar\omega)\}
\,.
\end{eqnarray}
Hence
\begin{eqnarray}
P(\lambda,\mu)&=&\left\{{(\lambda,\mu)-(\mu,\lambda)\over{\alpha^2
}}\right\}
(\omega-\bar\omega)
\ln\left({1-\omega\over{1-\bar\omega}}\right)\nonumber\\
&=&{\pi\over{\sqrt 3}}\left\{{(\lambda,\mu)-(\mu,\lambda)\over{\alpha^2
}}\right\}\,.
\label{n752}
\end{eqnarray}
(Note that the expression for $P(\lambda,\mu)$ is independent of $A$.)

Choosing $A$ so that $S(\lambda,\mu)$ is real gives
\begin{equation}
A=3\sqrt 3\,.
\end{equation}
The phase $P(\lambda,\mu)$ of \reg{n752} has already been found in
\cite{Lepowsky,LepWil}
(in the case $\alpha^2=3$),
but
only in the case of the locality of the twisted vertex operators corresponding
to pure momentum states. The above calculation verifies the intertwining
relation (and hence the representation property, by the above arguments) for
arbitrary (untwisted) states. Note that the work of \cite{Lepowsky,LepWil}
is for twists of
arbitrary order, and below we shall consider the generalisation of our approach
to twists of an arbitrary prime order.
\subsection{Self-duality and the definition of $\alpha$}
We now digress for a while from consideration of the locality
relations to discuss the self-duality of the resulting conformal field
theory.

Let us consider what condition it is necessary to impose on the
Eisenstein lattice in order to produce a self-dual $\ze_3$-twisted
conformal field theory, by which we mean that the partition function
of the theory,
\begin{equation}
\chi_\Hil(\tau)=\hbox{\rm Tr}_\Hil q^{L_0-c/24}\,,
\end{equation}
is invariant under the modular transformation $S:\tau\mapsto -1/\tau$
(and hence up to a phase under a general modular transformation). (See
\cite{PGmer} for a fuller discussion of the problem of satisfactorily
defining self-duality.)

There are two clear choices, to either make $\Lambda_R/\alpha$
self-dual (so
that the untwisted theory before projection is self-dual
\cite{DGMtwisted}) or to make $\Lambda$ (the ``natural'' structure
underlying this construction) self-dual. The first of these is clearly
the correct one, as it is in the orbifold spirit of starting with a
self-dual theory, projecting out by an automorphism and then adding in
twisted representations of the remaining theory to restore
self-duality (although this does rely on naive physical ideas of how
the various twisted sector contributions to the partition function and
the characters of the automorphism transform into one another under
modular transformations -- see {\em e.g.} \cite{Tuite:moon,Ginsparg}). [Note
that the two choices are in general inequivalent, since
\begin{equation}
\label{detrel}
\hbox{\rm det}
(\Lambda_R)
=3^d\hbox{\rm det} \Lambda\,,
\end{equation}
and so they are only simultaneously
self-dual for $\alpha=3^{1\over 4}$.]

Let us for the sake of completeness
show that self-duality of $\Lambda$ is inconsistent with self-duality
of the corresponding twisted conformal field theory. Concentrate on the
case $d=12$ ({\em i.e.} $c=24$), and let us evaluate the first few
terms in the expansion of the partition function for the theory obtained
by projecting out on to the $\theta=1$ space $\Hil_0\oplus\Hil_1\oplus
\Hil_2$. We would expect to obtain a partition function of the form
$j(\tau)+{\rm constant}$ (since $j$ is the unique modular invariant
function up to an additive constant
with a single simple pole at the origin in $q=e^{2\pi
i\tau}$).
Taking $\Lambda$ to be
self-dual fixes its theta function \cite{ConSlo}
\begin{equation}
\Theta_\Lambda(\tau)\in\ce[\phi_0(\tau /2),
\Delta_6(\tau)]\,,
\end{equation}
where
\begin{equation}
\phi_0(\tau /2)=1+6q+6q^3+6q^4+12q^7+\ldots
\end{equation}
\begin{equation}
\Delta_6(\tau)=q\prod_{n=1}^\infty(1-q^n)^6(1-q^{3n})^6\,,
\end{equation}
{\em i.e.}
\begin{equation}
\Theta_\Lambda(\tau)=\phi_0(\tau /2)^{12}+C_1\phi_0(\tau /2)^6
\Delta_6(\tau)+C_2\Delta_6(\tau)^2\,,
\end{equation}
where $C_1$, $C_2\in\ce$.
Equating the partition function of the theory
to $j(\tau)$ up to an additive constant and considering the first few terms in
the series expansion determines the degeneracy of the twisted sector ground
states (we assume the same value for the degeneracy in both sectors),
and we find this to
be non-integral, providing the required contradiction.

A similar calculation using $\Lambda_R/\alpha$ self-dual instead
appears to produce a self-dual partition function (the first few
coefficients are correct) provided the twisted sector ground state
degeneracy takes the
value 729 (=$3^6$, which is highly plausible!).

Let us therefore investigate the twisted sector degeneracy by studying the
representations of the twisted sector cocycles. Since $\Lambda$ is integral,
$(\lambda,\mu)\in\Epsilon$. Suppose $(\lambda,\mu)=m+n\omega$. Then
$P(\lambda,\mu)=\pi in/\alpha^2$, and $\langle\lambda,\mu\rangle=
2m-n$. Hence
\begin{equation}
\label{n758}
\gamma_\lambda\gamma_\mu=e^{2\pi im/\alpha^2}\gamma_\mu\gamma_\lambda
\,,
\end{equation}
using \reg{gammaT}.

But we have set $\det\left(\Lambda_R/\alpha\right)=1$.
Hence, by \reg{detrel}, $\det\Lambda=\left(\alpha^4/3\right)^d$. Since
$\Lambda$ is integral, $\det\Lambda\geq 1$, and so we must have $\alpha^2
\geq\sqrt 3$. Consider the set of integers $m_\lambda=(\lambda,\lambda)/
\alpha^2$, $\lambda\in\Lambda$. Suppose that their greatest common divisor
is 1. (This is certainly true for $\Lambda_R/\alpha$ a Niemeier lattice,
since $m_\lambda$ actually takes the value 1 for all but the Leech lattice,
and the result clearly holds in that case by inspection.) We know that
$(\lambda,\lambda)=M_\lambda\in\ze$, since $\Lambda$ is integral. This
gives us $\alpha^2=M_\lambda/m_\lambda\in\ze$. Also, we have, from the
argument following equation \reg{n29}, $\gamma_\lambda=\gamma_{\omega\lambda}$,
so that identifying the phases in \reg{n758} for $\lambda$ replaced by
$\omega\lambda$ we obtain the restriction $3m\in\alpha^2\ze$. If the
lattice $\Lambda$ is extremal, in the sense that $\Lambda_\beta=
\Lambda/\beta$ does not satisfy $(\lambda,\lambda)\in n\ze$ for all
$\lambda\in\Lambda_\beta$ and some (fixed) integer $n$ and factor $\beta>1$,
then
this is clearly equivalent to the statement that our values of $m$ above
have no common factor, and so the constraint arising from the invariance
of $\gamma_\lambda$ under the action of the lattice automorphism gives us
$\alpha^2=3$. We shall use this value from now on.
\subsection{Generalisation to Higher Order Symmetries}
Before continuing with our consideration of locality for the
$\ze_3$-twisted theory, let us note that the generalisation to higher
order symmetries $\ze_p$ is essentially obvious. (We shall take $p>3$
to be prime for simplicity, although there is no {\em a priori} reason
why we must do so.) No new features arise beyond those appearing in
the transition form the $\ze_2$ to the $\ze_3$ case.

We introduce $p-1$ sets of oscillators $b_n^{ki}$, $1\leq k\leq p-1$,
$1\leq i\leq d$, $n\in\ze$, and $p-1$ sets of twisted oscillators $c_r^{ki}$,
$1\leq k\leq p-1$, $1\leq i\leq d$, $r\in\ze/p-\ze$. These satisfy the
relations
\begin{equation}
[b_m^{ki},b_n^{lj}]=m\delta^{ij}\delta_{m,-n}\delta_{k+l,p}
\end{equation}
\begin{equation}
[c_r^{ki},c_s^{lj}]=r\delta^{ij}\delta_{r,-s}\delta_{kl}\,,
\end{equation}
and all other commutation relations vanish, with
$b_n^{ki\dagger}=b_{-n}^{p-k,i}$, $c_r^{ki\dagger}=c_{-r}^{ki}$. The straight
sector, $\Hil_0$, is created by the action of the $b$-oscillators on a set of
orthonormal states $|\lambda\rangle$ as usual, where $p^{ki}\equiv b_0^{ki}$
satisfy $p^{ki}|\lambda\rangle=\lambda_k|\lambda\rangle$ with $\lambda_k$ a
$d$-dimensional complex vector and $\bar{\lambda_k}=\lambda_{p-k}$
(any scaling, such as that introduced by the factor $\alpha$ in the
$\ze_3$ case, has been absorbed into the definition of $\lambda_k$),
while
we have $p-1$ twisted sectors, $\Hil_1,\ldots,\Hil_{p-1}$, where $\Hil_k$ is
created
by the action of the oscillators $c_{-r}^{ki}$, $r>0$, on some appropriate
twisted cocycle representation space. The twisted sectors are to form
meromorphic representations of the conformal field theory described by the
subspace of the twisted sector left invariant under the action of an
automorphism $\theta$ of order $p$.

The vertex operators in $\Hil_0$ are given by
\begin{equation}
\label{n770}
V_0(\psi,z)=:\prod_{k=1}^{p-1}\prod_{a=1}^{M_k}{i\over{(n_{ka}-1)!}}{d^{n_{ka
}}\over{dz^{n_{ka}}}}X^{ki_{ka}}(z)\cdot \exp
\left\{\sum_{k=1}^{p-1}i\lambda_{p-k}\cdot
X^k(z)\right\}:\sigma_\lambda\,,
\end{equation}
for
\begin{equation}
\label{n771}
\psi=\prod_{k=1}^{p-1}\prod_{a=1}^{M_k}b_{-n_{ka}}^{i_{ka}}|\lambda\rangle\,.
\end{equation}
$\sigma_\lambda$ is a cocycle operator, and
\begin{equation}
X^{ki}(z)=q^{ki}-ip^{ki}\ln z+i\sum_{n\neq 0}{b_n^{ki}\over
n}z^{-n}\,,
\end{equation}
with
\begin{equation}
\exp\left\{\sum_{k=1}^{p-1}\lambda_k\cdot
q^k\right\}|\mu\rangle=|\lambda+\mu\rangle\,,
\end{equation}
(addition inside the ket corresponding to addition of the corresponding
eigenvalues of the operators $p^{ki}$), and the obvious normal ordering.

Set $\hat\sigma_\lambda=\exp\left\{\sum_{k=1}^{p-1}\lambda_k\cdot
q^k\right\}\sigma_\lambda$. Then these operators must obey
\begin{equation}
\hat\sigma_\lambda\hat\sigma_\mu=(-1)^{\langle\lambda,\mu\rangle}\hat
\sigma_\mu\hat\sigma_\lambda\,,
\end{equation}
where the real symmetric bilinear form $\langle\cdot,\cdot\rangle$ is given by
\begin{equation}
\label{n775}
\langle\lambda,\mu\rangle\equiv\sum_{k=1}^{p-1}\lambda_k\cdot\mu_{p-k}\,,
\end{equation}
in order that the $V_0$ be local. This demonstrates the general feature of the
calculations for higher values of $p$. The operators in the untwisted sector
pair up ({\em i.e.} the $k$'th set with the $(p-k)$'th set) and we essentially
have $(p-1)/2$ separate commuting
calculations to perform, leading to the $(p-1)/2$
separate terms in \reg{n775} of the form $\lambda_k\cdot\mu_{p-k}+
\lambda_{p-k}\cdot\mu_k$. The case $p=2$ is distinct, as we have one set of
oscillators only, which pairs with itself. This is the origin of the new
feature noted above in the symmetry factors for the twisted operator cocycles.

The Virasoro state
\begin{equation}
\psi_L=\half\sum_{k=1}^{p-1}b_{-1}^kb_{-1}^{p-k}|0\rangle\,,
\end{equation}
and the moments of the corresponding vertex operator
\begin{equation}
L_n=\half:\sum_{k=1}^{p-1}\sum_{m=-\infty}^\infty b_m^k\cdot
b_{n-m}^{p-k}:\,,
\end{equation}
satisfy the Virasoro algebra with $c=(p-1)d$. The conformal weight of the state
\reg{n771} is given by
\begin{eqnarray}
h_\psi&=&\half\sum_{k=1}^{p-1}\lambda_k\cdot\lambda_{p-k}+\sum_{k=1}^{p-1}\sum_{
a=1}^{M_k}n_{ka}\nonumber\\
&=&\half\langle\lambda,\lambda\rangle+\sum_{k=1}^{p-1}\sum_{
a=1}^{M_k}n_{ka}\,.
\end{eqnarray}
Hence, the theory will be meromorphic provided $\langle\cdot,\cdot\rangle$
defines an even $(p-1)d$-dimensional $\ze$-lattice. (Clearly, we expect the
theory on $\Hil_0$ before projection to be self-dual if this lattice is
self-dual, {\em i.e.} $(p-1)d$ must be a multiple of 8.)

Define
\begin{equation}
\label{n779}
\theta b_n^{ki}\theta^{-1}={\omega_p}^kb_n^{ki}\,.
\end{equation}
{}From \reg{n770} we see
that, for $\theta$ to be an automorphism of the theory on $\Hil_0$, we must
have
\begin{equation}
\theta\lambda_k={\omega_p}^{-k}\lambda_k\,.
\end{equation}
The operators
\begin{equation}
\label{n781}
L_{nk}=\half:\sum_{l=1}^{p-1}\sum_{r\in\ze+{l\over p}}c_r^k\cdot c_{n-r}^k:
+\delta_{n0}{(p^2-1)d\over{24p}}
\end{equation}
satisfy the Virasoro algebra with $c=d(p-1)$. We assume for the moment that
these are the modes of the relevant twisted sector vertex operator. Hence, the
conformal weight of the twisted sector ground state is ${(p^2-1)d\over{24p}}$.
This must lie in $\ze/p$ in order that we may ultimately obtain a meromorphic
representation. Hence, we require $24\big| (p^2-1)d$. As remarked above, for
$\Hil_0$ to be a self-dual theory, $(p-1)d\in 8\ze$. Say $(p-1)d=8N$. Then we
require $3\big| (p+1)N$. So $c=(p-1)d$ must be a multiple of 24 if $3\not{
\big|}
(p+1)$, otherwise it must only be a multiple of 8.

Set
\begin{equation}
\label{n782a}
\theta c_{-r}^{ki}\theta^{-1}={\omega_p}^{\sigma_r}c_{-r}^{ki}
\end{equation}
\begin{equation}
\theta\chi_k={\omega_p}^{\rho_k}\chi_k\,,
\end{equation}
where $\chi_k$ is an element of the ground state of the twisted sector
$\Hil_k$,
$1\leq k\leq p-1$, and $\sigma_r$, $\rho_k\in\ze$. In order that the projection
on to $\theta=1$ states provides a meromorphic theory, we require
\begin{equation}
\sigma_{r_1}+\ldots+\sigma_{r_M}+\rho_k\in p\ze
\iff
r_1+\ldots +r_M+{(p^2-1)d\over{24p}}\in\ze\,,
\end{equation}
{\em i.e.} we take
\begin{equation}
\label{n784a}
\sigma_r=kpr
\end{equation}
\begin{equation}
\rho_k={kN(p+1)\over 3}\,.
\end{equation}
The vertex operators $V_k(\psi,z)$ in the sector $\Hil_k$ corresponding to a
state $\psi\in\Hil_0$ are defined as before, {\em i.e.} by writing down a naive
form and then correcting it by considering the $L_{-1}$ commutation relation.
Note that this introduces ${p-1\over 2}$ arbitrary constants, and these can be
chosen so that the additional contributions to $\epsilon(\lambda,\mu)$ in the
twisted sector are pure phases, as we did above, by choosing the coefficients
of the $p^{k'}\cdot p^{p-k}$ and $p^{p-k'}\cdot p^k$ terms to be equal and
opposite. Set
\begin{equation}
C^k_l(z)=i\sum_{r\in\ze+{l\over p}}{c^k_r\over r}z^{-r}\,.
\end{equation}
The couplings between the $b$-oscillators and the $C^k_l(z)$ in
$V_k(\psi,z)$ are given by requiring $\theta$ to be an automorphism, {\em i.e.}
suppose $b^l\leftrightarrow C^k_{\mu_k(l)}(z)$. Then $\theta$ is an
automorphism, from \reg{n779} and \reg{n782a},
if $l=\sigma_{-\mu_k(l)/p}$, {\em
i.e.} $l=-k\mu_k(l)$ from \reg{n784a}. Thus $\mu_k(l)=-lk^{p-2}$ (using
$k^{p-1}\equiv 1\bmod p$ for $k\not\equiv 0\bmod p$). We may proceed as before,
and check the representation property of the twisted sectors by verifying the
intertwining locality relation. Note that we need only check the relation
for one twisted sector, and
also that we need
merely to check the relevant analogues of (\ref{n748a}-\ref{n748c}). This
decomposes into ${p-1\over 2}$ copies of these relations, which then reduce to
(\ref{n750a}-\ref{n750d})
with ${1\over 3}$, ${2\over 3}$ replaced by ${k\over p}$, ${p-k\over
p}$ for $k=1,\ldots,{p-1\over 2}$. We recover the cocycle phase found in
\cite{Lepowsky,LepWil},
but, as before, we have extended the proof of locality to the full
Hilbert space, rather then just the momentum
states.

In the investigation below
of the remaining locality relations, we shall consider the case $p=3$ for
simplicity. But, as we have seen in the above, the generalisation to higher
values of $p$ is trivial (only the case $p=2$ is special).
\subsection{Further locality checks}
\subsubsection{Twisted Sector Conjugation}
The next locality relation in turn, {\em i.e.} \reg{six}, is analogous
to the relation
\begin{equation}
\Wbar(\chi,z)W(\rho,w)=\Wbar(\rho,w)W(\chi,z)
\end{equation}
in the $\ze_2$ case \cite{DGMtwisted}, in which it was found to be
equivalent
to a reality condition on the representation -- a relation which the
conjugation map on the twisted sector had to be chosen to satisfy.
An identical argument holds here, and we find that the locality
relation is equivalent to
\begin{equation}
(f_{\chi_1\phi\chi_2})^\ast=(-1)^{h_1+h_\phi+h_2}f_{\overline\chi_1
\overline\phi
\overline\chi_2}\,,
\label{n790}
\end{equation}
where
\begin{equation}
f_{\chi_1\phi\chi_2}=\langle\overline\chi_1|V_i(\phi,1)|\chi_2\rangle\,,
\end{equation}
with either $i=1$, $\chi_1\in\Hil_1$, $\chi_2\in\Hil_2$ or $i=2$,
$\chi_1\in\Hil_2$,
$\chi_2\in\Hil_1$, and $L_0\phi=h_\phi\phi$, $L_0\chi_j=h_j\chi_j$ for $j=1$,
2.
We see that if we take
\begin{equation}
\overline\chi=(-1)^{L_0}I\chi\,,
\end{equation}
for $\chi\in\Hil_1\cup\Hil_2$ with
\begin{equation}
Ic_rI^{-1}=-\bar c_r\,;\qquad
I\bar c_rI^{-1}=-c_r\,;\qquad
I\chi_0=M{\chi_0}^\ast\,,\end{equation}
for $\chi_0$ a twisted sector ground state, where
\begin{equation}
M^\dagger\gamma_\lambda M={\gamma_\lambda}^\ast\,,
\end{equation}
with $M$ a symmetric unitary matrix, then \reg{n790} is satisfied.
\subsubsection{Definition of $W_3(\chi,z)$}
In the case of intertwiners between twisted sectors, one cannot use
our previous trick of using action on the vacuum and the creation
axiom. We must instead take a more indirect approach.

Assuming that $W_3(\chi,z)$ satisfies the intertwining relation
\reg{seven} and we have the duality relation for the operators
$\V(\rho,z)$ (which must hold if they are to define a consistent
conformal field theory), {\em i.e.}
\begin{equation}
\V(\rho,z)\V(\phi,w)=\V\left(\V(\rho,z-w)\phi,w\right)
\end{equation}
({\em c.f.} \reg{duality}), then
\begin{eqnarray}
W_3(V_1(\phi_1,w_1)\chi_0,z)V_1(\phi_2,w_2)\chi_0&=&
V_2(\phi_2,w_2)W_3(V_1(\phi_1,w_1)\chi_0,z)\chi_0\nonumber\\
&=&V_2(\phi_2,w_2)V_2(\phi_1,w_1+z)W_3(\chi_0,z)\chi_0\,,
\label{n798}
\end{eqnarray}
where $\chi_0\in\Hil_1$ and $\phi_1$, $\phi_2\in\Hil_0$. Since $\Hil_1$ is
irreducible as a representation space for the theory described by the untwisted
sector, then we may take \reg{n798} as the definition of $W_3(\chi,z)$ in terms
of
an unknown state $F_{\chi_0}(z)\equiv W_3(\chi_0,z)\chi_0\in\Hil_2$ with
$\chi_0$
a (fixed) element of the twisted sector ground state.

Note that $\chi_0$ lies in the projected theory only when $c$ is a multiple of
72, but we expect this to make little difference to the arguments, {\em i.e.}
we
would expect a verification of the remaining locality relations for $c$ a
multiple of 72 to extend to other values of $c$ without any problem.

The definition \reg{n798} does not give an explicit form for $W_3(\chi,z)$.
Indeed,
constructing an explicit form once $F_{\chi_0}(z)$ is known is non-trivial,
since
it involves inverting the action of $V_1(\psi,z)$ on $\Hil_1$.
One of the advantages of our approach in
\cite{DGMtwisted,DGMtrialsumm,DGMtriality}
over that of FLM is
that it gives explicit forms for many operators whose properties FLM were
forced to infer by other considerations. This is here to some extent lost.
In addition, we would expect, from the symmetry of the nature of the
construction, that $\Wbar_3(\chi,z)$ would be of a similar form to
$W_3(\chi,z)$,
simply with the $c$- and $\bar c$-oscillators interchanged. This
symmetry would
be apparent if an explicit form for $W_3(\chi,z)$ were given, but is obscured
by
our choice of definition (although in Appendix \ref{App1} we do provide a proof
in the presence of the full locality relations of the theory, {\em i.e.}
assuming the conformal field theory to be consistent).
Still, in the absence of a natural definition ({\em
c.f.} the definition of $V_1$, $V_2$ by analogy with $V_0$), this appears to be
the only course of action to take.

The intertwining relation \reg{seven} for $W_3$ is now trivial, since
\begin{eqnarray}
&&W_3(V_1(\phi_1,w_1)\chi_0,z)V_1(\psi,\zeta)V_1(\phi_2,w_2)\chi_0\nonumber\\
&&\hskip100pt=W_3(V_1(\phi_1,w_1)\chi_0,z)V_1(V_0(\psi,\zeta-w_2)\phi_2,w_2)\chi_0
\nonumber\\
&&\hskip120pt\hbt{by the representation property for $\Hil_1$,}\nonumber\\
&&\hskip100pt=V_2(V_0(\psi,\zeta-w_2)\phi_2,w_2)V_2(\phi_1,w_1+z)F_{\chi_0}(z)\nonumber\\
&&\hskip100pt=V_2(\psi,\zeta)V_2(\phi_2,w_2)V_2(\phi_1,w_1+z)F_{\chi_0}(z)
\nonumber\\
&&\hskip120pt\hbt{by the representation property for $\Hil_2$,}\nonumber\\
&&\hskip100pt=V_2(\psi,\zeta)W_3(V_1(\phi_1,w_1)\chi_0,z)V_1(\phi_2,w_2)\chi_0\,,
\end{eqnarray}
and so we deduce that
\begin{equation}
W_3(V_1(\phi_1,w_1)\chi_0,z)V_1(\psi,\zeta)=V_2(\psi,\zeta)
W_3(V_1(\phi_1,w_1)\chi_0,z)\,,
\end{equation}
as required.
\subsubsection{Evaluation of $F_{\chi_0}(z)$}
In order to verify the locality relation \reg{eight}, {\em i.e.}
\begin{equation}
W_3(\chi,z)W_1(\rho,\zeta)=W_3(\rho,\zeta)W_1(\chi,z)\,,
\label{n7102}
\end{equation}
we need only check its action on the vacuum $|0\rangle$, since we have the
intertwining relations for both $W_3$ and $W_1$. Then, by the action of $W_1$
on $|0\rangle$, {\em c.f.} \reg{n46}, we see that \reg{n7102} is equivalent to
the
``skew-symmetry'' relation
\begin{equation}
e^{zL_{-1}}W_3(\chi,-z)\rho=W_3(\rho,z)\chi\,,
\end{equation}
{\em c.f.} \reg{W1}. Putting $\chi=V_1(\phi_1,w_1)\chi_0$ and
$\rho=V_1(\phi_2,w_2)\chi_0$, we see from \reg{n798} that this is equivalent to
the same relation for $F_{\chi_0}(z)$, {\em i.e.}
\begin{equation}
F_{\chi_0}(z)=e^{zL_{-1}}F_{\chi_0}(-z)\,.
\label{n7104}
\end{equation}
[Note that, if we had chosen the two ground states in the definition of
$F_{\chi_0}(z)\equiv W_3(\chi_0,z)\chi_0$ to be distinct, then they would be
interchanged between the left and the right hand sides of \reg{n7104}. The
choice
of coincident ground states has been made for simplicity.] Note that
the definition of
$W_1$ in terms of $V_1$ was forced upon us by a skew-symmetry relation.
However, in this case, skew-symmetry cannot be used to define $W_3$, but merely
to give a condition which must be satisfied by the arbitrary quantity appearing
in its definition.

Consider the locality relation \reg{nine}. We see from a discussion
mirroring that in \cite{DGMtwisted} that we need
only verify the
relation
\begin{equation}
\langle\chi_0|\Wbar_3(\overline\chi_0,z)V_2(\phi,w)W_3(\chi_0,\zeta)|\chi_0\rangle
=\langle\chi_0|W_1(\chi_0,\zeta)V_0(\phi,w)\Wbar_1(\overline\chi_0,z)|\chi_0\rangle
\,,\end{equation}
where $\phi\in\Hil_0$ is arbitrary (with $\theta=1$).
This reduces to
\begin{equation}
\langle F_{\chi_0}(z)|V_2(\phi,w)|F_{\chi_0}(\zeta)\rangle=
\langle G_{\chi_0}(1/\zeta^\ast)|V_0(\phi,w)|G_{\chi_0}
(1/z^\ast)\rangle\,,
\label{n7110}
\end{equation}
where
\begin{eqnarray}
G_{\chi_0}(z)&\equiv&W_1(\chi_0,1/z^\ast)^\dagger\chi_0\nonumber\\
&=&\sum_{\lambda\in\Lambda}{\chi_0}^\dagger{\gamma_\lambda}^\dagger\chi_0
e^{A_1(1/z^\ast)^\dagger}|\lambda\rangle\,.
\end{eqnarray}
This is analogous to the final relation in the $\ze_2$ case, and is
discussed in the next section.

We also have the locality relation \reg{ten}. Now
\begin{eqnarray}
W_3(\chi,z)\Wbar_3(\phi,w)W_3(\chi_0,\alpha)\chi_0&=&
W_3(\chi,z)W_1(\chi_0,\alpha)\Wbar_1(\phi,w)\chi_0\hb{\hfill by
\reg{nine}}\nonumber\\
&=&W_3(\chi_0,\alpha)W_1(\chi,z)\Wbar_1(\phi,w)\chi_0\hb{\hfill by
\reg{eight}}\nonumber\\
&=&W_2(\chi,z)\Wbar_2(\phi,w)W_3(\chi_0,\alpha)\chi_0
\end{eqnarray}
by \reg{seven}, since
\begin{equation}
W_1(\chi,z)\Wbar_1(\phi,w)=V_1(\Wbar_2(\chi,z-w)\phi,w)
\end{equation}
and
\begin{equation}
W_2(\chi,z)\Wbar_2(\phi,w)=V_2(\Wbar_2(\chi,z-w)\phi,w)\,,
\end{equation}
by duality for $\V(\chi,z)$ (using relations independent of this one!). So, we
deduce that \reg{ten} is valid when acting on the state $F_{\chi_0}(z)$, and
hence for all states by the interwining relations.

To conclude, all that remains to do in order to prove that this construction
defines a consistent (meromorphic bosonic) conformal field theory is to check
\reg{twelve} and define $F_{\chi_0}(z)\in\Hil_2$ such that \reg{n7104},
\reg{eleven} and
\reg{n7110} hold. The approach which we will adopt is to use
\reg{n7104} and \reg{eleven} to
give restrictions on the definition of $F_{\chi_0}(z)$, and then the
idea is to use any postulated expression satisfying these to check \reg{n7110}.
(\reg{twelve} may be checked independently of all this.) The remainder of this
sub-section is devoted to finding a plausible expression for
$F_{\chi_0}(z)$, and in the next sub-section, we consider the relations
\reg{n7110}
and \reg{twelve}. It should be noted that
both the definition of $F_{\chi_0}(z)$ and the completion of the
consistency check for the conformal field theory in the next sub-section are as
yet
incomplete.

First, note that
in order to check \reg{eleven}, we need only verify
\begin{equation}
\Wbar_2(\chi_0,z)F_{\chi_0}(\zeta)=\Wbar_2(\chi_0,\zeta)F_{\chi_0}(z)\,.
\label{n7115}
\end{equation}
We initially postulate that $W_3(\chi_0,z)\chi_0$ is of the following form
\begin{eqnarray}
\label{n7116}
F_{\chi_0}(z)&=&z^{-\Delta_0}\exp\left\{\sum_{r,s>0}\bar c_{-r}\cdot\bar c_{-s}
z^{r+s}D_{rs}\right\}\chi(\chi_0)\nonumber\\
&=&z^{-\Delta_0}e^{D(z)}\chi(\chi_0)\,,
\end{eqnarray}
where $\chi(\chi_0)$ is a twisted sector ground state of conformal weight
$\Delta_0$. Note that the relations \reg{n7104} and \reg{n7115} provide no
restrictions
on $\chi(\chi_0)$. It will be determined by the locality relation \reg{n7110}.
[The factor $z^{-\Delta_0}$ is needed to ensure the correct
commutation relation with $L_0$. This is also to be expected from the
postulated symmetry between $W_3$ and $\Wbar_3$, since the factor of
$z^{-2\Delta_0}$ in the definition of $\Wbar_3$ would then also lead to an
overall factor of $z^{-2\Delta_0}(1/z)^{-\Delta_0}=z^{-\Delta_0}$ in that
operator.]

Let us assume that
\begin{equation}
e^{{1\over z}L_1}F_{\chi_0}(\zeta)=\left(1-{\zeta\over z}
\right)^{-2\Delta_0}F_{\chi_0}\left({z\zeta\over{z-\zeta}}\right)\,,
\label{n7117}
\end{equation}
which must hold if we are to have locality for the full
theory (see \cite{PGmer} for a discussion of the M\"obius
transformation properties of vertex operators corresponding to primary
states).
Then \reg{n7115}
reduces to invariance under the interchange of $z$ and $\zeta$ of the
expression
\begin{eqnarray}
\Wbar_2(\chi_0,z)F_{\chi_0}(\zeta)&=&
(z-\zeta)^{-2\Delta_0}\sum_{\lambda\in\Lambda}e^{A_2(1/z^\ast)^\dagger}\langle
\overline\chi_0|{\gamma_\lambda}^\dagger e^{B_2^-(1/z^\ast)^\dagger}
F_{\chi_0}\left({z\zeta\over{z-\zeta}}\right)|\lambda\rangle\nonumber\\
&=&(z-\zeta)^{-2\Delta_0}\left({z\zeta\over{z-\zeta}}\right)^{-\Delta_0}
\sum_{\lambda\in\Lambda}e^{A_2(1/z^\ast)^\dagger}{\overline\chi_0}^\dagger
{\gamma_\lambda}^\dagger\chi(\chi_0)\nonumber\\
&&\qquad\cdot\langle\widetilde 0|e^{B_2^-(1/z^\ast)^\dagger}e^{D(z\zeta/
(z-\zeta))}|\widetilde 0
\rangle|\lambda\rangle\,,
\label{n7118}
\end{eqnarray}
where $|\widetilde 0\rangle$ denotes a ``vacuum'' state in $\Hil_2$, {\em i.e.}
$\langle\widetilde 0|\widetilde 0\rangle=1$, $\bar c_t|\widetilde 0\rangle=0$
for $t>0$.
Since $\Delta_0$ is a multiple of 4 (for $c$ a multiple of 72),
$(z-\zeta)^{-\Delta_0}$ is clearly invariant under the interchange. So, we are
left
with the expression
\begin{equation}
e^{A_2(1/z^\ast)^\dagger}\langle\widetilde 0|e^{B_2^-(1/z^\ast)^\dagger}
e^{D(z\zeta/(z-\zeta))}|\widetilde 0
\rangle|\lambda\rangle\,.
\label{n7119}
\end{equation}
Now, an obvious solution to the skew-symmetry relation \reg{n7104} is to
take
\begin{equation}
F_{\chi_0}(z)=z^{-\Delta_0}e^{\half zL_{-1}}\chi(\chi_0)\,.
\label{n7120}
\end{equation}
However, this is easily seen not to satisfy \reg{n7117}. This observation
curtails a naive hope that invariance of \reg{n7119} under $z\leftrightarrow
\zeta$ (under which the argument of $D$ changes sign) could be
deduced simply as a consequence of the relation \reg{n7104}, which
details the effect of such a sign change.

The matrix element $\langle\widetilde 0|\ldots|\widetilde 0\rangle$ gives
\begin{equation}
\exp\left\{\sum_{m,n\geq 0\atop{r,s>0}}rB^-_{mr}\left({1\over z}
\right)sB^-_{ns}\left({
1\over z}\right)
\left({z\zeta\over{z-\zeta}}\right)^{r+s}D_{rs}b_{-m}\cdot\bar b_{-n}
\right\}\,.
\label{n7121}
\end{equation}
So, \reg{n7119} becomes
\begin{equation}
\exp\left\{\sum_{m,n\geq 0}P_{mn}(\zeta,z)b_{-m}\cdot\bar b_{-n}\right\}\,,
\label{n7122}
\end{equation}
where
\begin{equation}
P_{mn}(\zeta,z)=\sum_{r,s>0}D_{rs}\left({r\atop m}\right)\left({s\atop
n}\right)
\left({\zeta\over{\zeta-z}}\right)^{r+s}(-z)^{m+n}+A^1_{mn}\left({1\over z}
\right)\,.
\label{n7123}
\end{equation}
Hence
we require invariance under the interchange $z\leftrightarrow \zeta$ of the
expression
\begin{eqnarray}
P(x,y,\zeta,z)&\equiv&\sum_{m,n\geq 0}P_{mn}(\zeta,z)x^my^n\nonumber\\
&=&\sum_{m,n\geq 0}A^1_{mn}\left({1\over z}
\right)x^my^n+\sum_{r,s>0}D_{rs}\left({(1-
zx)\zeta\over{\zeta-z}}\right)^r\left({(1-
zy)\zeta\over{\zeta-z}}\right)^s\,.
\label{n7124}
\end{eqnarray}
We have an expression for the first of these two sums from \reg{n750a}. Denote
the
second sum by $g\left({(1-
zx)\zeta\over{\zeta-z}},{(1-
zy)\zeta\over{\zeta-z}}\right)$.

Then we require
\begin{equation}
g(a,b)-g(1-a,1-b)=-(1-\omega)\ln\left({
(1-a)^{1\over 3}-\omega(1-b)^{1\over 3}\over
{(-a)^{1\over 3}-\omega(-b)^{1\over 3}}}\right)+(\omega\rightarrow
\bar\omega)\,.
\label{n7125}
\end{equation}

Now consider the constraint given by the skew-symmetry
relation \reg{n7104} on $F_{\chi_0}(z)$.
In order to consider this, we require a Fock space representation of
$e^{zL_{-1}}$. This is derived in Appendix \ref{App2}.
Set $d_r=\bar c_r/\sqrt{|r|}$ and $e_s=\bar c_s/\sqrt{|s|}$. Then we have, in
matrix notation,
\begin{equation}
e^{zL_{-1}}=e^{d^\dagger\bar A(z)e^\dagger}
:e^{d^\dagger\bar B(z)d}:\quad
:e^{e^\dagger\bar C(z)e}:\,,
\label{n7126}
\end{equation}
where
\begin{equation}\bar A_{rs}(z)\equiv\sqrt{rs}A_{rs}z^{r+s}\end{equation}
\begin{equation}\bar B_{rr'}(z)\equiv\sqrt{rr'}B_{rr'}z^{r-r'}\end{equation}
\begin{equation}\bar
C_{ss'}(z)\equiv\sqrt{ss'}C_{ss'}z^{s-s'}\,.\end{equation}
Using the simple normal ordering relation
\begin{equation}:e^{\alpha^\dagger G\alpha}:
e^{\beta \alpha^\dagger}=e^{\alpha^\dagger (1+G)\beta}
:e^{\alpha^\dagger G\alpha}:\,,\end{equation}
where
$\alpha_n$ are bosonic creation and annihilation operators with commutation
relations normalised to unity ({\em c.f.} the $d$- and $e$-oscillators above),
we
see that
\begin{equation}e^{zL_{-1}}F_{\chi_0}(-z)=z^{-\Delta_0}e^{d^\dagger\bar
A(z)e^\dagger}
:e^{d^\dagger\bar B(z)d}:\quad
:e^{e^\dagger\bar C(z)e}:
e^{d^\dagger\bar D(z)e^\dagger}\chi(\chi_0)\,,\end{equation}
where
\begin{equation}\bar D_{rs}(z)\equiv\sqrt{rs}D_{rs}(-z)^{r+s}\,,\end{equation}
becomes
\begin{equation}
\label{n7131}
z^{-\Delta_0}e^{d^\dagger\bar A(z)e^\dagger}
e^{d^\dagger(1+\bar B(z))\bar D(z)(1+\bar C(z))^Te^\dagger}
\chi(\chi_0)\,.\end{equation}
Therefore, \reg{n7104} is satisfied if this is equal to
\begin{equation}
\label{n7132}
F_{\chi_0}(z)\equiv z^{-\Delta_0}e^{d^\dagger\bar D(-z)e^\dagger}
\chi(\chi_0)\,.\end{equation}
We replace $d_{-r}$ by $x^{-r}/\sqrt r$ and $e_{-s}$ by $y^{-s}/\sqrt s$ for
$r$, $s>0$ in \reg{n7131} and \reg{n7132}, and the problem then reduces to
verifying
the equality of the generating functions for the coefficients of the
oscillators in the exponentials.

{}From \reg{n7132}, we obtain the generating function
\begin{eqnarray}
f(x,y,z)&=&\sum_{r,s>0}x^{-r}y^{-s}D_{rs}z^{r+s}\nonumber\\
&\equiv&g\left({z\over x},{z\over y}\right)\,.
\end{eqnarray}
The generating function which arises from the second term of \reg{n7131} is
given
by
\begin{eqnarray}
h(x,y,z)&=&\sum_{r,s>0\atop{r',s'>0}}{x^{-r}y^{-s}\over{\sqrt{rs}}}(\delta_{rr'}+
\bar B_{rr'}(z))(\delta_{ss'}+
\bar C_{ss'}(z))\sqrt{r's'}D_{r's'}(-z)^{r'+s'}\nonumber\\
&=&\sum_{r',s'>0\atop{r>r',s>s'}}{x^{-r}y^{-s}\over{\sqrt{rs}}}\sqrt{r's'}
\left(\delta_{rr'}+\sqrt{rr'}{(-z)^{r-r'}\over{r'}}\left({-r'\atop
{r-r'}}\right)\right)\nonumber\\
&&\qquad\cdot\left(\delta_{ss'}+\sqrt{ss'}{(-z)^{s-s'}\over{s'}}\left({-s'\atop
{s-s'}}\right)\right)
D_{r's'}(-z)^{r'+s'}\nonumber\\
&=&\sum_{r',s'>0\atop{n,m\geq 0}}x^{-m-r'}y^{-n-s'}z^{m+n}(-1)^{m+n}
\left({-r'\atop m}\right)\left({-s'\atop
n}\right)D_{r's'}(-z)^{r'+s'}\nonumber\\
&=&\sum_{r',s'>0}D_{r's'}\left({z\over{x-z}}\right)^{r'}
\left({z\over{y-z}}\right)^{s'}\nonumber\\
&\equiv&g\left({z\over{z-x}},{z\over{z-y}}\right)\,.
\end{eqnarray}
The first term of \reg{n7131} contributes
\begin{equation}k(x,y,z)\equiv\sum_{r,s>0}x^{-r}y^{-s}A_{rs}z^{r+s}\,.\end{equation}
So, \reg{n7104} is satisfied if
\begin{equation}g\left({z\over x},{z\over y}\right)\equiv
g\left({z\over{z-x}},{z\over{z-y}}\right)+k(x,y,z)\,.\end{equation}
Using \reg{appendix}, we can easily check that
\begin{equation}
\label{n7137}
k(x,y,z)=-(1-\omega)\ln\left({
(1-y/z)^{1\over 3}-\omega(1-x/z)^{1\over 3}
\over
{
(-y/z)^{1\over 3}-\omega(-x/z)^{1\over 3}
}
}\right)+(\omega\rightarrow
\bar\omega)
\,,\end{equation}
and hence skew-symmetry is equivalent to
\begin{eqnarray}
\label{n7138}
g(a,b)-g\left({a\over{a-1}},{b\over{b-1}}\right)&=&-(1-\omega)
\ln\left({(1-1/b)^{1\over 3}-\omega(1-1/a)^{1\over 3}\over
{(1/b)^{1\over 3}-\omega(1/a)^{1\over 3}}}\right)\nonumber\\
&&\qquad+(\omega\rightarrow
\bar\omega)\,.\end{eqnarray}
Now, it is easily verified that the relation \reg{n7117} is
equivalent to
\begin{equation}
\label{n7139}
L_1P_n=(n-1+\Delta_0)P_{n-1}\,,\end{equation}
where
\begin{equation}F_{\chi_0}(z)=z^{-\Delta_0}\sum_{n\geq
0}P_nz^n\,,\end{equation}
and hence to
\begin{equation}
\label{n7141}
(r+s)D_{r s}=(r+1)D_{r+1 s}+(s+1)D_{r s+1}+
{\textstyle{2\over 9}}D_{{1\over 3}
s}D_{r {2\over 3}}\,.\end{equation}
Note that this set of relations is sufficient to determine the
coefficients $D_{rs}$ at level $n$ ({\em i.e.} $r+s=n$)
in terms of those at level $n-1$ together with one indeterminate.
This single unknown at each level could be provided by consideration
of \reg{n7125} and \reg{n7138} for $a=b$ to produce an expression for $g(a,a)$.
However, we note that in our argument of Appendix \ref{App1} the function
\begin{equation}h(a)\equiv\left.{d\over{da}}{d\over{db}}g(a,b)\right|_{a=b}
\end{equation}
appeared in the expression for a particular correlation function, and so
we feel that this is a more natural object to consider, not least because
it removes the logarithms from \reg{n7125} and \reg{n7138}, which give
\begin{equation}h(a)-h(1-a)={2a-1\over{9a^2(1-a)^2}}\end{equation}
and
\begin{equation}h(a)-{1\over{(1-a)^4}}h\left({a\over{a-1}}\right)={2-a\over{9a(1-a)^2}}
\,.\end{equation}
Noting the obvious solution to these, we write
\begin{equation}h(a)={1\over{9a(1-a)^2}}+k(a)\,,
\label{n7145}
\end{equation}
where
\begin{equation}
\label{n7146}
k(a)=k(1-a)={1\over{(1-a)^4}}k\left({a\over{a-1}}\right)\,.\end{equation}
Now, from $g(a,b)\equiv\sum_{r,s>0}D_{rs}a^rb^s$, we see that $h(a)$ has
a simple pole at the origin with residue ${2\over 9}D_{{1\over 3} {2\over
3}}$. But the skew-symmetry relation \reg{n7104} implies that
$D_{{1\over 3} {2\over
3}}={1\over 2}$, and so we see from \reg{n7145} that $k(a)$ is regular at
$0$. From \reg{n7146}, $k(a)$ is regular also at $1$ and $\infty$,
and since these are the only possible poles if our conformal field theory
is to be consistent
(using the fact that $h(a)$
is essentially a correlation function), we see, from
Liouville's theorem, that $k(a)$ is constant.
Further, \reg{n7146} shows us that this constant value is zero.

Now we may use the relations \reg{n7141} to evaluate the coefficients
$D_{rs}$ at successive levels. Carrying through this calculation for
the first few levels, we arrive at the (surprisingly asymmetric) conjecture
\begin{equation}
\label{n7147}
D_{rs}={(-1)^{r+s}\over{r+s}}\left({1\over{3s}}-1\right)
\left({-{2\over 3}\atop{r-{1\over 3}}}\right)
\left({-{1\over 3}\atop{s-{2\over 3}}}\right)\,.\end{equation}
Note that these coefficients have been independently derived in
\cite{Gato:tvos}
by rather different techniques.
We must now check that this satisfies \reg{n7141}, \reg{n7125} and \reg{n7138}.
The relation \reg{n7141} is trivially verified. Evaluating $g(a,b)$
directly however is difficult. Let us therefore consider instead
the derivative with respect to $b$ of the other two relations.
We have
\begin{eqnarray}
\mu(a,b)\equiv{d\over{db}}g(a,b)&=&-a^{-{1\over 3}}b^{1\over 3}
{d\over{db}}\sum_{r,s>0}b^ra^s{(-1)^{r+s}\over{r+s}}
\left({-{1\over 3}\atop{r-{1\over 3}}}\right)
\left({-{2\over 3}\atop{s-{2\over 3}}}\right)
\nonumber\\
&=&{\left({a\over b}\right)^{1\over 3}-\left({a(a-1)\over{b(b-1)}}
\right)^{1\over 3}\over{a-b}}\,.
\end{eqnarray}
We may then easily check that this satisfies the $b$ derivative
of \reg{n7125} and \reg{n7138}, hence verifying these relations up to addition
of an arbitrary function of $a$. Noting that both sides of \reg{n7138}
vanish at $b=0$ shows that this arbitrary function vanishes identically
in this case. However, we may have a non-zero anomaly in \reg{n7125}.
Since we have already noted that direct calculation of $g(a,b)$
seems difficult, let us consider
\begin{eqnarray}
\label{n7149}
\rho(a,b)\equiv{d\over{da}}g(a,b)&=&\sum_{r,s>0}{a^{r-1}b^s\over{3s}}
(-1)^{r+s}\left({-{2\over 3}\atop{r-{1\over 3}}}\right)
\left({-{1\over 3}\atop{s-{2\over 3}}}\right)\nonumber\\
&&-
a^{-{1\over 3}}{d\over{da}}\sum_{r,s>0}{(-1)^{r+s}\over{r+s}}
a^{r+{1\over 3}}b^s
\left({-{2\over 3}\atop{r-{1\over 3}}}\right)
\left({-{1\over 3}\atop{s-{2\over 3}}}\right)\,.
\end{eqnarray}
We may evaluate the second term using \reg{n7137}.
Substituting this into the derivative of \reg{n7125} we obtain the appropriate
term on the right hand side. However, we also have a contribution from
the first term of equation \reg{n7149}. The $b$ derivative of this term
is $-b^{-{1\over 3}}(1-b)^{-{1\over 3}}
a^{-{2\over 3}}(1-a)^{-{2\over 3}}/3$, and so the anomalous piece
on the right hand side of \reg{n7125} is
\begin{equation}
\label{n7150}
f(a)=-{1\over 3}\int_0^1{db\over{b^{1\over 3}(1-b)^{1\over 3}}}
\int^a_{1\over 2}{dx\over{x^{2\over 3}(1-x)^{2\over 3}}}\,,
\end{equation}
(noting that $f({1\over 2})=0$ by considering \reg{n7125} for
$a=b={1\over 2}$).

We must now adjust our definition \reg{n7116} of $F_{\chi_0}(z)$ in order
to compensate for this, without affecting either \reg{n7138} or \reg{n7139}.
Let us therefore try
\begin{eqnarray}
F_{\chi_0}(z)&=&z^{-\Delta_0}\sum_{\alpha\in\Gamma}\exp\left\{
\sum_{r>0}{\bar\alpha\cdot\bar c_{-r}\over{\sqrt 3}}F_rz^r\right\}
\exp\left\{
\sum_{r,s>0}\bar c_{-r}\cdot\bar c_{-s}D_{r,s}z^{r+s}\right\}
\chi_\alpha(\chi_0)\nonumber\\
&\equiv&z^{-\Delta_0}\sum_{\alpha\in\Gamma}e^{F_\alpha(z)}e^{D(z)}
\chi_\alpha(\chi_0)
\,,
\label{n7151}
\end{eqnarray}
where $\chi_\alpha(\chi_0)$ is a twisted sector ground state and
$\Gamma$ is an Eisenstein lattice containing $\Lambda$
(and the $\sqrt 3$ arises naturally from the definition of the action
of $b_0$ on $|\lambda\rangle$).
Equation \reg{n7139} then determines the coefficients
\begin{equation}
\label{n7152}
F_r={F\over r}\left({-{2\over 3}\atop{r-{1\over 3}}}\right)
(-1)^{r-{1\over 3}}\,,
\end{equation}
for some constant $F$. To check that this does not affect skew-symmetry,
we need to go through a similar argument to that leading up to \reg{n7138}.
The $a$ derivative of the contribution of \reg{n7152} to the generating
function is clearly $\hat g(a)=a^{-{2\over 3}}(1-a)^{-{2\over 3}}$.
Hence, to avoid contradicting \reg{n7138}, we could have
\begin{equation}\hat g(a)=-{1\over{(1-a)^2}}\hat g\left({a\over{a-1}}\right)\,.
\label{n7153}
\end{equation}
This is clearly not true. However, note that we must take the sum over
$\alpha\in\Gamma$ into account.
\reg{n7153} is valid up to a factor of $-\omega$, and so if we can arrange
so that only every sixth term in the expansion of the first exponential
in \reg{n7151} actually contributes, then we have the desired result.
Since $\Gamma$ is an Eisenstein lattice, then for $\alpha\in\Gamma$ we
have $-\alpha$, $\omega\alpha$, $\bar\omega\alpha\in\Gamma$. So, provided
that
\begin{equation}\chi_\alpha(\chi_0)=\chi_{-\alpha}(\chi_0)=
\chi_{\omega\alpha}(\chi_0)\,,
\label{n7154}
\end{equation}
we see that, on summing over the lattice $\Gamma$, all but every sixth
term {\em does} vanish and so
skew-symmetry
remains valid.
[Note that $\chi_\alpha(\chi_0)=\chi_{\omega\alpha}(\chi_0)$
is in general necessary in any case in order that $F_{\chi_0}(z)$
be meromorphic.]

Finally, we must check that our modification cancels the anomaly $f(a)$.
There are two parts to this. Firstly, we verify that the contribution
of the $F_r$'s to the relevant generating function produces the
appropriate term. Secondly, we must check that the other contributions
from the sum over $\alpha\in\Gamma$ ({\em i.e.} for $\alpha\neq\lambda$)
cause no problems in \reg{n7118}.

So, let us begin by considering the term $\alpha=\lambda$. The additional
contribution to the generating function $g(a,b)$ due to our new
exponential term is given by
\begin{equation}\phi(x,z,\zeta)=\sum_{r>0\atop{n\geq
0}}F_r\left({z\zeta\over{z-\zeta}}
\right)^r\cdot r\cdot{(-z)^{n-r}\over r}\left({r\atop
n}\right)x^n\,,\end{equation}
from commuting $B_2^-(1/z^\ast)^\dagger$ past $e^{F_\alpha\left(z\zeta/
(z-\zeta)\right)}$ in the matrix element analogous to \reg{n7119}.
Rewriting this in terms of the variable $a=(1-zx)\zeta/(\zeta-z)$ as in
\reg{n7125}, we find the additional term
\begin{equation}
\label{n7156}
\phi(a)=\int_0^a{F db\over{(1-b)^{2\over 3}b^{2\over 3}}}\,.
\end{equation}
So, we can cancel our anomaly $f(a)$ if $\phi(a)-\phi(1-a)+f(a)=0$.
We see that this is achieved by setting
\begin{equation}
\label{n7157}
F={1\over 6}\int_0^1{db\over{b^{1\over 3}(1-b)^{1\over 3}}}=
{\Gamma\left({2\over 3}\right)^2\over{2\Gamma\left({1\over 3}\right)}}
\,,\end{equation}
using the well known identity for the beta function.

Now consider the remainder of the contribution of our additional
exponential. The $\alpha$-dependent piece of \reg{n7118} is
\begin{equation}\sum_{\alpha\in\Gamma}{\overline\chi_0}^\dagger{\gamma_\lambda}^\dagger
\chi_\alpha(\chi_0)\exp\left\{\sum_{n\geq 0}{\bar\alpha\cdot b_n\over{\sqrt 3}}
\phi_n\right\}\,,\end{equation}
where
\begin{equation}\phi(a)=\sum_{n\geq 0}\phi_nx^n\,.\end{equation}
This is
\begin{equation}\sum_{\alpha\in\Gamma}{\overline\chi_0}^\dagger{\gamma_\lambda}^\dagger
\chi_{\alpha+\lambda}(\chi_0)
\exp\left\{\sum_{n\geq 0}{\bar\alpha\cdot b_n\over{\sqrt 3}}
\phi_n\right\}
\exp\left\{\sum_{n\geq 0}{\bar\lambda\cdot b_n\over{\sqrt 3}}
\phi_n\right\}
\,,
\label{n7160}
\end{equation}
where the final term is that which we have used to cancel our anomaly.
In order that the remainder of \reg{n7160} be invariant under the locality
transformation $a\mapsto 1-a$, we require
\begin{equation}\sum_{\alpha\in\gamma}{\overline\chi_0}^\dagger{\gamma_\lambda}^\dagger
\chi_{\alpha+\lambda}(\chi_0)e^{
\bar\alpha\cdot\phi(a)/\sqrt 3}
\end{equation}
be invariant, abusing our notation slightly and writing $\phi$ as a
vector. But
\begin{equation}\phi(a)=\phi(1)-\phi(1-a)\,,\end{equation}
from \reg{n7156}. Hence, we must have
\begin{equation}\sum_{\alpha\in\Gamma}{\overline\chi_0}^\dagger{\gamma_\lambda}^\dagger
\chi_{\alpha+\lambda}(\chi_0)e^{\bar\alpha\cdot\lambda
\phi(1)/3}e^{\bar\alpha\cdot\phi(a)/\sqrt 3}
=\sum_{\alpha\in\Gamma}{\overline\chi_0}^\dagger{\gamma_\lambda}^\dagger
\chi_{-\alpha+\lambda}(\chi_0)e^{
-\bar\alpha\cdot\phi(a)/\sqrt 3}
\,.\end{equation}
Now
\begin{equation}\phi(1)=F\int_0^1{da\over{a^{2\over 3}(1-a)^{2\over 3}}}=
\hhalf\Gamma\left({\textstyle{1\over 3}}\right)
\Gamma\left({\textstyle{2\over 3}}\right)={\pi\over{\sqrt 3}}\,,
\end{equation}
using the beta function identity once more, together with \reg{n7157}.
Let us try
\begin{equation}\chi_\alpha(\chi_0)=\eta_\alpha\chi(\chi_0)\,,\end{equation}
where $\eta_\alpha$ is a scalar.
Remembering the restriction \reg{n7154} on $\chi_\alpha$, we try
\begin{equation}\eta_\alpha=e^{\epsilon\alpha\cdot\bar\alpha}\,,\end{equation}
for some (complex) scalar $\epsilon$, and try to set
\begin{equation}
\label{n7167}
\eta_{\alpha+\lambda}e^{{\bar\alpha\cdot\lambda\over 3}\phi(1)}=
\eta_{-\alpha+\lambda}\,.\end{equation}
Considering the real part of this gives
${\rm Re}\,\epsilon=-{\pi\over{12\sqrt 3}}$, leaving us with a phase
$e^{(\bar\alpha\cdot\lambda-\alpha\cdot\bar\lambda)\pi/6\sqrt 3}$.
[Note that this
negative value for ${\rm Re}\,\epsilon$ guarantees convergence of
our expression \reg{n7151} for $F_{\chi_0}(z)$, which was not well-defined
{\em a priori}.]

Now, the overall factor of ${\overline\chi_0}^\dagger{\gamma_\lambda}^\dagger
\chi(\chi_0)$ could at best be arranged so that it vanishes except
when $\lambda\in(1-\omega)\Lambda\equiv\Lambda_0$, or at least some
coset of $\Lambda_0\in\Lambda$. [Note that for $\Lambda_0$ the twisted
cocycle symmetry factors vanish and so $\gamma_\lambda$ is a multiple
of the identity in an irreducible representation (which we have seen
that we must have from our earlier considerations of self-duality.]
Suppose $\lambda_0\in\Lambda_0$. Say $\lambda_0=(1-\omega)\lambda$, with
$\lambda\in\Lambda$ and set $\bar\alpha\cdot\lambda=m+n\omega$ (assuming
$\Gamma\subset\Lambda^\ast$). Then $\bar\alpha\cdot\lambda_0-\alpha\cdot
\bar\lambda_0=\sqrt 3i(2n-m)$, so that our remaining phase to be corrected
in \reg{n7167} is $e^{\pi i(2n-m)/6}$. Choosing ${\rm Im}\,\epsilon=\pi/12$
(and hence $\epsilon=\pi\omega/6\sqrt 3$), we find that we are left with a
phase
$e^{\pi i(m+n)/3}$. Now our condition on $\Lambda$, {\em i.e.} that
$(\lambda,\lambda)\in3\ze$ for all $\lambda\in\Lambda$, gives us,
assuming $\Gamma=\Lambda$, $m+n\in 3\ze$, and hence our residual phase
is simply $(-1)^{(m+n)/3}$.
However, specific examples show that $(m+n)/3$ can be odd.

We do not yet understand what further modifications can be made to the
definition of $F_{\chi_0}(z)$ to correct for this, though note that,
if we do not require a self-dual theory, we may project $\Lambda$ on to
a sub-lattice $\Lambda'$ such that $\lambda\cdot\bar\lambda\in 6\ze$ for
all $\lambda\in\Lambda'$,
for which the locality relation then holds exactly.
However, we would expect the construction to
work in the self-dual case also, by analogy with the lattice constructions
described later. [Note that we can easily insert gamma matrices into
the definition of $\chi_\alpha(\chi_0)$ which potentially correct the
phase. However, as we find when we consider the locality relations
$W_1(\chi,z)\Wbar_2(\rho,\zeta)=W_1(\rho,\zeta)\Wbar_2(\chi,z)$ and
$\Wbar_3(\chi,z)W_3(\rho,\zeta)=W_1(\rho,\zeta)\Wbar_1(\chi,z)$, the optimal
solution would seem to be to double the value of $F$ ({\em c.f.}
\reg{n7157}), and so we leave this investigation until such matters are
fully understood.]

Also note that the expression for $F_{\chi_0}(z)$ satisfying the relations
which we have examined so far is not unique. In particular, we are
free to insert an overall scale (even zero!) into our definition. This
scale in particular, and presumably the full expression, will be fixed
(at least up to a trivial ``gauge" symmetry of the ground states) by
a full examination of the locality relation
$\Wbar_3(\chi,z)W_3(\rho,\zeta)=W_1(\rho,\zeta)\Wbar_1(\chi,z)$.

The locality relations
$W_1(\chi,z)\Wbar_2(\rho,\zeta)=W_1(\rho,\zeta)\Wbar_2(\chi,z)$ and
$\Wbar_3(\chi,z)W_3(\rho,\zeta)=W_1(\rho,\zeta)\Wbar_1(\chi,z)$ are
the analogues of the final relation in the $\ze_2$ case. The approach
to their verification follows essentially along the lines set out in
\cite{DGMtwisted}. Much of this has been done, though the details and
one or two problems remain to be completed, and so we leave this to a
future publication.

We now look at the possible results of such a construction, and
analogies to the links with coding theory discovered in the $\ze_2$
case in \cite{DGMtrialsumm,DGMtriality}.
\subsection{Results of the Construction for $c=24$}
As noted in section \ref{lattice}, an
$\Epsilon$-lattice has a natural $\ze_3$ symmetry, which is
inherited by the corresponding $\ze$-lattice $\Lambda_R/\alpha$.
[The exact form of the converse statement, {\em i.e.} that a $\ze$-lattice
with a third order NFPA can be written in the form $\Lambda_R/\alpha$ for
some $\alpha$, where $\Lambda$ is an integral $\Epsilon$-lattice, is
not immediately apparent, although we shall find it unnecessary, at least
in the 24-dimensional case upon which we are focusing, to consider this.]
So, we must check which of the Niemeier
lattices (the even self-dual lattices in 24 dimensions)
admits a third order NFPA. As shown by Venkov \cite{Venkov}, each of the
lattices is uniquely determined by its set of minimal vectors (those
of length squared two), which form the
root system of a simply-laced semi-simple Lie algebra. A third order NFPA of a
Niemeier lattice must map the minimal vectors into themselves, and so provides
a third order NFPA of the root lattice of the corresponding semi-simple Lie
algebra. Clearly, each component of the algebra (if the algebra is non-simple)
must map into itself under the action of the automorphism (otherwise
the components must be permuted and we could
find a non-zero
vector fixed by this action). The third order NFPA's of the root
lattices of simple Lie algebras have been classified \cite{Mythesis}
and only $A_2$,
$D_4$, $E_6$, $E_8$, $F_4$ and $G_2$ admit such automorphisms. (In each case
there is only one appropriate conjugacy class in the automorphism group.) So,
from
the list of possibilities derived by
Venkov, we see that the only lattices which could possibly admit a third
order NFPA are ${E_8}^3$, ${D_4}^6$, ${A_2}^{12}$, ${E_6}^4$ and
$\Lambda_{24}$. Note that, in the case of ${E_8}^3$, the lattice {\em is} the
root lattice of the Lie algebra, and so it {\em does} admit the symmetry. But,
in the remaining cases, we must check that the symmetry can be extended to the
full lattice.

The twist-invariant subalgebras corresponding to the first four possibilities
are ${A_8}^3$, ${A_2}^6$, $U(1)^{24}$ and ${A_2}^{12}$ respectively. Since the
conformal weight of the twisted sector ground state is ${4\over 3}$, then the
states of lowest conformal weight coming from the twisted sector are of weight
2, and so there is no enhancement of this (the algebra is generated by
the zero modes of the vertex operators corresponding to the states of
conformal weight one \cite{PGmer}), {\it i.e.} the algebra
of the new theory $\widehat\Hil(\Lambda_R/\alpha)$ is just the
twist-invariant subalgebra of
that of $\Hil(\Lambda_R/\alpha)$ (which is simply the Lie algebra
corresponding to the Niemeier lattice).
We prove a theorem in \cite{DGMtriality} that when the rank of the Lie
algebra corresponding to a conformal field theory is equal to the
central charge, the conformal field theory is isomorphic to
$\Hil(\Sigma)$ for some even lattice $\Sigma$. Thus, the theories in
the first,
third and fourth cases must just be $\Hil({A_8}^3)$, $\Hil(\Lambda_{24})$
and $\Hil({A_2}^{12})$ respectively, if they do exist ({\em i.e.} if the
$\ze_3$ construction is consistent).
Only $\widehat\Hil({D_4}^6)$ and $\widehat\Hil(\Lambda_{24})$ could
provide new theories.

If ${D_4}^6$ admits a third order NFPA (and $\widehat\Hil({D_4}^6)$ is
consistent as
a conformal field theory), then this would definitely be a new theory, since
the algebra ${A_2}^6$ does not occur in the list of results in
\cite{DGMtrialsumm}.
If the
conjecture of FLM, that the self-dual conformal field theory with no states of
weight one and with least value of $c$ is unique, is correct, then we must have
$\widehat\Hil(\Lambda_{24})=V^\natural$($=\widetilde\Hil(\Lambda_{24})$,
the $\ze_2$-twisted orbifold theory),
{\em i.e.} we
would have a new construction of the natural Monster module.
We shall discuss this further later.
It is hoped that such a
construction could throw more light upon the triality structure which, although
perhaps rendered a little less mysterious as a result of earlier work
\cite{DGMtriality,DGMtrialsumm},
we feel still needs to be explained in a more satisfactory setting.

We could check explicitly whether the postulated $\ze_3$ symmetries
extend to the Niemeier lattices. However, we shall obtain this
result naturally from
the work of the next sub-section in all but the case of
the lattice corresponding
to ${D_4}^6$. Hence,
we now merely check whether the third order NFPA of the ${D_4}^6$
root lattice extends
to the Niemeier lattice ${D_4}^6$.

We may describe a Niemeier lattice $\Sigma$
by specifying the appropriate Lie algebra and a set of glue vectors
\cite{ConSlo}.
Let $\Sigma_M$ be the root lattice of the Lie algebra, {\it i.e.} the lattice
spanned by the minimal vectors of $\Sigma$. Suppose $\Sigma_M=\Sigma_1\oplus
\ldots\oplus\Sigma_k$, where the $\Sigma_i$ are the root lattices of the
simple components of the Lie algebra. Then any $\lambda\in\Sigma$ must be an
element of the dual of $\Sigma$ (note that we only require that $\Sigma$ be
integral), {\it i.e.} $\lambda=(\lambda_1,\ldots,\lambda_k)$, where each
$\lambda_i\in{\Sigma_i}^\ast$. Clearly we need only consider cosets of
$\Sigma_i$ in ${\Sigma_i}^\ast$. A standard system of representatives of
these are referred to as the glue vectors of $\Sigma_i$ (usually chosen to be
of minimal length). The lattice $\Sigma$ is then generated by
$\Sigma_M$ together with a set of glue vectors
$(\lambda_1,\ldots,\lambda_k)$, with each $\lambda_i$ a glue vector for
$\Sigma_i$. We say that the components $\Sigma_1,\ldots,\Sigma_k$ have been
glued together by the glue vectors. (Note that this description can be
applied for any
integral lattice.)

Let us take as generator matrix for the root lattice of $D_4$ the
matrix
\begin{equation}
\left(
\begin{array}{cccc}
2&0&0&0\\
1&1&0&0\\
1&0&1&0\\
1&0&0&1
\end{array}
\right)
\end{equation}
Then a set of glue vectors is given by
\begin{equation}
\label{gluea}
[0]=(0,0,0,0)\,;\qquad
[1]=(\hhalf,\hhalf,\hhalf,\hhalf)
\end{equation}
\begin{equation}
\label{glueb}
[2]=(0,0,0,1)\,;\qquad
[3]=(\hhalf,\hhalf,\hhalf,-\hhalf)\,.
\end{equation}
The matrix
\begin{equation}
\left(
\begin{array}{cccc}
-1&1&1&1\cr
-1&-1&-1&1\cr
-1&1&-1&-1\cr
-1&-1&1&-1
\end{array}
\right)
\end{equation}
describes a third order automorphism of this lattice \cite{Mythesis}.

A set of glue vectors for the Niemeier lattice ${D_4}^6$ is given, using the
notation of \reg{gluea} and \reg{glueb},
by $[111111]$ and $[0(02332)]$, where the parentheses
include that we should also include all cyclic permutations. Then, in order
to check that the symmetry $M$ does not extend to the Niemeier
lattice,
we need only
check that $M$ does not map all of the glue vectors into elements of
the
lattice.
In fact, we must check that no element in the same conjugacy class of the
automorphism group of the root lattice of $D_4$ as $M$ maps the glue
vectors into the Niemeier lattice. Consider $M'=h^{-1}Mh$, for $h$ an
arbitrary element of this automorphism group. Denote by $[n]_\ast$ the
coset containing $[n]$ in ${D_4}^\ast$ modulo $D_4$. Then, considering
the glue corresponding to the first $D_4$ component of the Niemeier
lattice ${D_4}^6$, we see that we must have $M'[1]_\ast=[0]_\ast$ or
$M'[1]_\ast=[1]_\ast$ if $M'$ is to extend to the full lattice.
Since $[0]_\ast=D_4$ and $\left(M'\right)^{-1}$
maps $D_4$ into itself, we see that we must have the second of the two
possibilities,
{\em i.e.} $Mh[1]_\ast=h[1]_\ast$. However, we can easily check that
$M[n]_\ast=[n]_\ast$ only for $n=0$. Hence $h[1]_\ast=[0]_\ast$, and
acting with $h^{-1}$ on this we find that $h^{-1}[0]_\ast=[1]_\ast$, in
contradiction to the fact that $h^{-1}\in\Aut(D_4)$.
[Note that in the case of the other Niemeier lattices which potentially
admit a third order NFPA, and which we shall show in the next
sub-section
in fact do,
we should in principle consider the possibility that the single conjugacy
class in the automorphism group of the corresponding root lattice
splits into more than one in the automorphism group of the full lattice.
However, we have seen that the conformal field theories which are produced
must be independent of these potentially distinct $\ze_3$ twists, since
they are of the form $\Hil(\Lambda)$, determined solely by the
twist-invariant subalgebra arising from the root lattice, at least in the case
of all but the Leech lattice, and so we need not consider this situation
for our purposes.]
Hence ${D_4}^6$ does not admit a third order NFPA, and the potentially
new theory with algebra ${A_2}^6$ does not exist.

What we have shown is that we obtain at most four
theories (we still must check that ${A_2}^{12}$, $\Lambda_{24}$
and ${E_6}^4$ do admit third
order NFPA's), three of which would be merely theories of the form
$\Hil(\Lambda)$ for
$\Lambda$ another Niemeier lattice, while the fourth is most likely to be the
Monster module.
\subsection{Straight and Twisted Constructions of $\Epsilon$-Lattices
from $\ \ \ \ \ $ Ternary Codes}
Considering the results of the previous sub-section, we see that
the possibilities
are as shown in figure \ref{fig1}, where upward-sloping arrows indicate the
usual
FKS construction
and downward-sloping arrows the $\ze_3$-twisted construction.
\setlength{\unitlength}{1cm}
\begin{figure}[htb]
\begin{center}
\begin{picture}(5,14)
\put(0.7,13.65){lattice}
\put(3.75,13.65){cft}
\put(0.9,10.7){${E_8}^3$}
\put(3.75,11.9){$\Hil({E_8}^3)$}
\put(3.75,9.5){$\Hil({A_8}^3)$}
\put(1.9,10.85){\vector(3,2){1.7}}
\put(1.9,10.85){\vector(3,-2){1.7}}
\put(0.9,6.4){${E_6}^4$}
\put(0.9,4){${A_2}^{12}$}
\put(0.9,1.6){$\Lambda_{24}$}
\put(3.75,7.6){$\Hil({E_6}^4)$}
\put(3.75,5.2){$\Hil({A_2}^{12})$}
\put(3.75,2.8){$\Hil(\Lambda_{24})$}
\put(3.75,0.4){$V^\natural$}
\put(1.9,6.55){\vector(3,2){1.7}}
\put(1.9,6.55){\vector(3,-2){1.7}}
\put(1.9,4.15){\vector(3,2){1.7}}
\put(1.9,4.15){\vector(3,-2){1.7}}
\put(1.9,1.75){\vector(3,2){1.7}}
\put(1.9,1.75){\vector(3,-2){1.7}}
\end{picture}
\end{center}
\caption{}
\label{fig1}
\end{figure}
This is very
reminiscent
of the work
on binary codes and $\ze_2$-twisted conformal field theories, and leads
us to strongly suspect both that these results are as shown in the figure
({\em i.e.} that the twists lift to the Niemeier lattices and the twisted
conformal field theories are consistent) and that there is some code
structure to the left of the figure.

However, there appears to be only space for two or perhaps three
codes to be inserted. Knowing there are only three self-dual
ternary codes of length 12 (two of which contain codewords of weight
12), we are led to consider a construction in 12 dimensions, {\em i.e.}
to consider a complex construction. Having realised in the previous
sections that it is in fact the 12 dimensional $\Epsilon$-lattice which
is the relevant structure to consider, it now seems natural to attempt
to construct $d$-dimensional Eisenstein lattices from length $d$
ternary codes, and then the corresponding $2d$-dimensional $\ze$-lattices
will carry a natural $\ze_3$ NFPA.

We shall generalise a pair of constructions due to Sloane
\cite{ConSlo,Sloane:codes}.

Let $\hat\C$ be a self-dual $d$-dimensional ternary code. Define
the ``straight" construction by
\begin{equation}\Lambda_\Epsilon(\hat\C)=\hat\C+(\omega-\bar\omega)
\Epsilon^d\,.
\end{equation}
This is clearly an $\Epsilon$-lattice (note that $\omega\hat c=\hat c-
(\omega-\bar\omega)\bar\omega\hat c$ for $\hat c\in\hat\C$, and so
we have an Eisenstein module) and further $(\lambda,\lambda)\in 3\ze$
for $\lambda\in\Lambda_\Epsilon(\hat\C)$. From the expression given
by Sloane, we see that $\det\left(\Lambda_\Epsilon(\hat\C)_R/\sqrt
3\right)=1$, and so $\Lambda_\Epsilon(\hat\C)_R/\sqrt
3$ is an even self-dual $\ze$-lattice with a third order NFPA induced
by the map $\lambda\mapsto\omega\lambda$ on $\Lambda_\Epsilon(\hat\C)$.
(Note that the dimension of a self-dual ternary code must be
a multiple of 4, and this is consistent with the constraint that
the dimension of an even self-dual ($\ze$-)lattice be a multiple of 8.)

For $\one\in\hat\C$ ($\one=(1,1,\ldots,1)$),
and hence $d$ a multiple of 12 (say $d=12D$),
define the ``twisted" construction by
\begin{equation}\tilde\Lambda_\Epsilon(\hat\C)=\Lambda_0(\hat\C)\cup
\Lambda_1(\hat\C)\cup\Lambda_2(\hat\C)\,,\end{equation}
where
\begin{equation}\Lambda_0(\hat\C)=\hat\C+(\omega-\bar\omega)\Epsilon^d_0
\end{equation}
\begin{equation}\Lambda_1(\hat\C)=\hat\C+(\omega-\bar\omega)\left(\Epsilon^d_D
+{\textstyle{1\over 3}}\one\right)
\end{equation}
\begin{equation}\Lambda_2(\hat\C)=\hat\C+(\omega-\bar\omega)\left(\Epsilon^d_{-D}
-{\textstyle{1\over 3}}\one\right)\,,
\end{equation}
with
\begin{equation}\Epsilon^d_\rho\equiv\left\{x=(x_1,\ldots,x_d)
\in\Epsilon^d\,:\,
\sum_{i=1}^dx_i\equiv{\rho\bmod (\omega-\bar\omega)}\right\}\,.\end{equation}
We can easily check that this is a lattice. (Note that the requirement
that it is an $\Epsilon$-module implies, as above, that $(\omega-\bar\omega)
\bar\omega\hat c\in(\omega-\bar\omega)\Epsilon^d_0$ for $\hat c\in
\hat\C$. This is found to be equivalent to $\one\cdot\hat c\in 3\ze$,
{\it i.e.} $\one\in\hat\C^\ast=\hat\C$.) We also find that $(\lambda,
\lambda)\in 3\ze$, and $\tilde\Lambda_\Epsilon(\hat\C)_R/\sqrt 3$
is an even self-dual $2d$-dimensional $\ze$-lattice with a $\ze_3$
NFPA.

[Note that these two constructions are a slightly generalised form of those
given in \cite{ConSlo}.
We have extended the twisted construction to work for $d$
an arbitrary multiple of 12, rather than just for $d\equiv{12\bmod 36}$, and
rescaled our definitions so that the corresponding real lattices are
even and self-dual.]

Let us consider the results of these two constructions for $d=12$.
Referring to \cite{MallPlessSloan},
we see that there are three self-dual codes of length 12,
{\it i.e.} $\C_{12}$, $4\C_3(12)$ and $3\C_4$. $3\C_4$ does not contain
$\one$, and so cannot induce a twisted construction. Let us identify the
corresponding Niemeier lattices.

The vectors of length squared 2 in $\Lambda_\Epsilon(\hat\C)_R/\sqrt 3$
correspond to vectors of norm 3 in $\Lambda_\Epsilon(\hat\C)$. These
are the $27n_3$ vectors of the form
\begin{equation}
\label{n766}
\hat c+(\omega-\bar\omega)x\,,\end{equation}
with $\hat c\in\hat\C$ of weight 3 and $x=(x_1,\ldots,x_d)\in\Epsilon^d$
with $x_i=\omega\hat c_i$, $-\bar\omega\hat c_i$ or 0, together with the
$6d$ vectors
\begin{equation}\pm(\omega-\bar\omega)\zeta e_i\,,\end{equation}
with $\zeta=1$, $\omega$ or $\bar\omega$ and $e_i$ the unit vector
containing 1 in the $i$'th coordinate. The corresponding vectors in
$\tilde\Lambda_\Epsilon(\hat\C)$ are simply the
$9n_3$ vectors of the form \reg{n766} with $\sum_{i=1}^dx_i={0\bmod (\omega-
\bar\omega)}$. Hence, the Coxeter number of the corresponding Niemeier
lattice is $9n_3/8+3$ in the straight case and $3n_3/8$ in the twisted
case. Substituting in the $n_3$ values of 0, 8 and 24 for $\C_{12}$,
$4\C_3(12)$ and $3\C_4$ respectively, we see that we can extend figure
\ref{fig1}
to produce the structure shown in figure \ref{fig2}.
\setlength{\unitlength}{1cm}
\begin{figure}[htb]
\begin{center}
\begin{picture}(8,14)
\put(3.7,13.65){lattice}
\put(6.75,13.65){cft}
\put(3.9,10.7){${E_8}^3$}
\put(6.75,11.9){$\Hil({E_8}^3)$}
\put(6.75,9.5){$\Hil({A_8}^3)$}
\put(0.9,10.7){$3\C_4$}
\put(0.7,13.65){code}
\put(1.75,10.85){\vector(1,0){2}}
\put(4.9,10.85){\vector(3,2){1.7}}
\put(4.9,10.85){\vector(3,-2){1.7}}
\put(3.9,6.4){${E_6}^4$}
\put(3.9,4){${A_2}^{12}$}
\put(0.4,5.2){$4\C_3(12)$}
\put(1.0,2.8){$\C_{12}$}
\put(1.9,5.35){\vector(3,2){1.7}}
\put(1.9,5.35){\vector(3,-2){1.7}}
\put(1.9,2.95){\vector(3,2){1.7}}
\put(1.9,2.95){\vector(3,-2){1.7}}
\put(3.9,1.6){$\Lambda_{24}$}
\put(6.75,7.6){$\Hil({E_6}^4)$}
\put(6.75,5.2){$\Hil({A_2}^{12})$}
\put(6.75,2.8){$\Hil(\Lambda_{24})$}
\put(6.75,0.4){$V^\natural$}
\put(4.9,6.55){\vector(3,2){1.7}}
\put(4.9,6.55){\vector(3,-2){1.7}}
\put(4.9,4.15){\vector(3,2){1.7}}
\put(4.9,4.15){\vector(3,-2){1.7}}
\put(4.9,1.75){\vector(3,2){1.7}}
\put(4.9,1.75){\vector(3,-2){1.7}}
\end{picture}
\end{center}
\caption{}
\label{fig2}
\end{figure}
(The Coxeter number
does not identify the lattices ${E_6}^4$ or ${E_8}^3$ uniquely (see
\cite{ConSlo}),
but clearly $3\C_4$ produces 3 orthogonal components under the straight
construction, while we may pick out ${E_6}^4$ with equal ease.)

This confirms that we do now indeed have the correct analogy. Note that
it also verifies without the need for additional work that the lattices
${E_6}^4$, ${A_2}^{12}$ and $\Lambda_{24}$ do admit an extension of the third
order NFPA's of the corresponding root lattices. It is interesting to
note that most of the possible $\ze_3$ NFPA's of the Niemeier lattices are
realised (the exception being ${D_4}^6$) and further that all of the ones
which are realised exist due to a link with ternary codes.
\subsection{``Triality'' Structure}
It has been demonstrated by Dong and Mason \cite{ZpOrbifold} that
$\widehat\Hil(\Lambda_\Epsilon(\C_{12}))$ is the Monster conformal
field theory $V^\natural$, by showing essentially that there is a
unique irreducible representation of the affine Griess algebra (formed
by the modes of the vertex operators corresponding to the states of
conformal weight two). However, there should also exist an analogous
proof to that in the $\ze_2$ case by FLM \cite{FLMbook} (and
generalised to a wider class of conformal field theories in
\cite{DGMtrialsumm,DGMtriality}).
First note that $\C_{12}$, the ternary Golay code, is characterised by
properties similar to that of the corresponding binary code, {\em i.e.}
it is the unique self-dual ternary code of smallest length containing
no codewords of weight 3.
The complex Leech lattice $\Lambda_\Epsilon(\C_{12})$ has symmetry group
$6\,Suz$ \cite{ConSlo}, where $Suz$ is the Suzuki group, one of the finite
sporadic simple groups. We note that the non-abelian composition factor
in the centraliser of an element in the conjugacy class $3-$ of the
Monster is $F_{3-}=Suz$.
Comparing this with the results described in
\cite{DGMtrialsumm,DGMtriality},
{\em i.e.}
$F_{2-}=Co_1=\Aut(\Lambda_{24})/\ze_2$, suggests that the automorphism
group of this conformal field theory is again the Monster.
An analogous analysis
to that given for the construction of the triality operator in
\cite{DGMtrialsumm,DGMtriality}
should complete the Suzuki group to the Monster, and also generalise to
codes other than $\C_{12}$. At the heart of that analysis lay the affine
$su(2)^{24}$ algebra, while in this case we see that we have an affine
$su(3)^{12}$ algebra, generated by the modes of the vertex operators
corresponding to the weight one states
\begin{equation}a^i_{-1}|0\rangle\,,\qquad \bar a^i_{-1}|0\rangle\,,\qquad
|\pm(\omega-\bar\omega)\zeta\rangle\,,\end{equation}
where $\zeta=1$, $\omega$, $\bar\omega$, for each $i=1,\ldots,12$.
However, this analysis remains to be carried through.
\subsection{Results for Higher Order Twists}
Finally, for the sake of completeness, we
look at the possible theories which we may
obtain from orbifolding with respect to higher order NFPA's.
Concentrate again on the $c=24$ theories.
We have $c=(p-1)d$. So $(p-1)\big| 24$, and the only primes $p\geq 3$ possible
are $p=3$, 5, 7, 13. The simple Lie algebras whose root systems admit such
symmetries are $A_{p-1}$ in each case, together with $E_8$ in the case $p=5$.
Hence, the Niemeier lattices which could possibly admit appropriate NFPA's are
as shown in figure \ref{fig3}.
\setlength{\unitlength}{1cm}
\begin{figure}[htbp]
\begin{center}
\begin{picture}(8,20)
\put(0.65,21.65){twist}
\put(3.7,21.65){lattice}
\put(6.75,21.65){cft}
\put(0.65,17){$p=5$}
\put(4,19.4){${E_8}^3$}
\put(4,17){${A_4}^6$}
\put(4,14.6){$\Lambda_{24}$}
\put(6.75,20.6){$\Hil({E_8}^3)$}
\put(6.75,18.2){$\Hil({A_4}^6)$}
\put(6.75,15.8){$\Hil(\Lambda_{24})$}
\put(6.75,13.4){$V^\natural$}
\put(4.9,19.55){\vector(3,2){1.7}}
\put(4.9,19.55){\vector(3,-2){1.7}}
\put(4.9,17.15){\vector(3,2){1.7}}
\put(4.9,17.15){\vector(3,-2){1.7}}
\put(4.9,14.75){\vector(3,2){1.7}}
\put(4.9,14.75){\vector(3,-2){1.7}}
\put(6.75,12){$\Hil({A_6}^4)$}
\put(6.75,9.6){$\Hil(\Lambda_{24})$}
\put(6.75,7.2){$V^\natural$}
\put(4,10.8){${A_6}^4$}
\put(4,8.4){$\Lambda_{24}$}
\put(0.65,9.6){$p=7$}
\put(4.9,10.95){\vector(3,2){1.7}}
\put(4.9,10.95){\vector(3,-2){1.7}}
\put(4.9,8.55){\vector(3,2){1.7}}
\put(4.9,8.55){\vector(3,-2){1.7}}
\put(6.75,5.5){$\Hil({A_{12}}^2)$}
\put(6.75,3.1){$\Hil(\Lambda_{24})$}
\put(6.75,0.7){$V^\natural$}
\put(4,4.3){${A_{12}}^2$}
\put(4,1.9){$\Lambda_{24}$}
\put(0.65,3.1){$p=13$}
\put(4.9,4.45){\vector(3,2){1.7}}
\put(4.9,4.45){\vector(3,-2){1.7}}
\put(4.9,2.05){\vector(3,2){1.7}}
\put(4.9,2.05){\vector(3,-2){1.7}}
\end{picture}
\end{center}
\caption{}
\label{fig3}
\end{figure}
(Note that ${E_8}^3$ certainly does.) The twist invariant subalgebra is
${A_4}^6$ in the case of ${E_8}^3$. So, as before, we see that this is not a
new theory. The remaining cases produce a self-dual conformal field theory with
either no $L_0=1$ states (in the case of $\Lambda_{24}$) or 24 such states (in
the case of ${A_{p-1}}^{24/(p-1)}$). The twist-invariant subalgebras in the
second case are $U(1)^{24}$, so that these theories are simply
$\Hil(\Lambda_{24})$ again. The theories in the first case should be
$V^\natural$
once more, by the FLM conjecture. Note that
we do not know of any constructions of lattices from codes over the
field ${\Bbb F}_p$
which
would complete the structure of figure \ref{fig3},
so that in order to verify explicitly
that the Niemeier lattices admit an extension of the $\ze_p$ NFPA
induced by their corresponding root lattice we should have to do some
more work.
(It is however known that the Leech lattice does possess such
automorphisms.)
\section{Conclusions}
We have defined explicitly and essentially verified the consistency of
$\ze_3$ NFPA orbifolds of lattice conformal field theories, as well as
indicating the corresponding results for higher prime ordered twists.

A link with ternary codes analogous to that of $\ze_2$ twists with
binary codes has been demonstrated, and it is to be hoped that this
will further illuminate the still somewhat mysterious structure of the
Monster, as well as generalising aspects of this structure to other
conformal field theories constructed in this way from ternary codes
other than the ternary Golay code.

The relationship to other work on ternary codes \cite{PSMcodes} is as
yet unclear, though promises to reward further investigation with
regard to the completion of the classification problem for the
self-dual $c=24$ conformal field theories \cite{PSMorb}.

Though patently much of the motivation for this work is as an abstract
generalisation of previous work, the physical applications of this
construction include, as described in the introduction, ``realistic''
string models containing twisted bosonic fields.
This general mechanism of increasing symmetry by combining
sectors is also
present in superstring models involving twisted world-sheet fermions.
Our techniques can clearly be extended to include fermionic fields,
and the more explicit results for the structure of the vertex operators
than previously found should, in all these models, aid in calculations of,
for example, string scattering amplitudes.

Also, the techniques used in the derivation of the implicit form of the twisted
sector-twisted sector intertwiner $W_3$, which form the essential core of
this paper, are exploited in \cite{PSMreps}
to provide a new perspective on representations of conformal field theories,
and perhaps consequently a new approach to other concepts such as fusion could
result.
\appendix
\section{Twisted Sector Representation of $e^{zL_{-1}}$
for Third Order Twist}
\label{App1}
In this appendix we derive an explicitly normal ordered expression for
$e^{zL_{-1}}$ in the sector $\Hil_2$ of the $\ze_3$-twisted theory.

Write
\begin{eqnarray}
\label{k1}
&&e^{zL_{-1}}=\exp\left\{\sum_{r,s>0}A_{rs}\bar c_{-r}\cdot\bar c_{-s}z^{r+s}
\right\}
:\exp\left\{\sum_{r,r'>0}B_{rr'}\bar c_{-r}\cdot\bar c_{r'}z^{r-r'}
\right\}\nonumber\\
&&\hskip150pt
\exp\left\{\sum_{s,s'>0}C_{ss'}\bar c_{-s}\cdot\bar c_{s'}z^{s-s'}
\right\}:\,.
\end{eqnarray}
Note the absence of a term bilinear in the annihilation operators, since we
require $\langle\widetilde 0|e^{zL_{-1}}=0$. Define
\begin{equation}
\label{k2}
\langle\widetilde x,\widetilde y|\equiv\langle\widetilde 0|\sum_{r,s>0}
{x^{-r}y^{-s}\over{rs}}\bar c_r\cdot\bar c_s\,.\end{equation}
Set
\begin{eqnarray}
\label{k3}
A(x,y,z)&&\equiv\langle\widetilde x,\widetilde y|e^{zL_{-1}}|\widetilde
0\rangle\nonumber\\
&=&\sum_{r,s>0}A_{rs}\left({z\over x}\right)^r\left({z\over y}\right)^sd\,,
\end{eqnarray}
from \reg{k1} and \reg{k2}.
But
\begin{eqnarray}
{d\over{dz}}A(x,y,z)&=&\langle\widetilde x,\widetilde y|e^{zL_{-1}}L_{-1}
|\widetilde
0\rangle\nonumber\\
&=&\langle\widetilde x,\widetilde y|e^{zL_{-1}}\bar c_{-{1\over 3}}\cdot\bar
c_{
-{2\over 3}}|\widetilde
0\rangle\nonumber\\
&=&\langle\widetilde x,\widetilde y|\sum_{n,m\geq 0}(-z)^{n+m}
\left({-{1\over 3}\atop n}\right)
\left({-{2\over 3}\atop m}\right)
\bar c_{-{1\over 3}-n}\cdot\bar c_{
-{2\over 3}-m}e^{zL_{-1}}|\widetilde
0\rangle\nonumber\\
&=&\langle\widetilde 0|\sum_{r,s>0}
{x^{-r}y^{-s}\over{rs}}\bar c_r\cdot\bar c_s
\sum_{n,m\geq 0}(-z)^{n+m}\nonumber\\
&&\qquad\qquad
\left({-{1\over 3}\atop n}\right)
\left({-{2\over 3}\atop m}\right)
\bar c_{-{1\over 3}-n}\cdot\bar c_{
-{2\over 3}-m}e^{zL_{-1}}|\widetilde
0\rangle\,.
\end{eqnarray}
Clearly, only the first term in the action of the series expansion of
$e^{zL_{-1}}$ on $|\widetilde 0\rangle$ gives a non-zero contribution
to
\reg{k3},
as we require an equal number of creation and annihilation operators.
Hence
\begin{eqnarray}
{d\over{dz}}A(x,y,z)&=&
\langle\widetilde 0|\sum_{r,s>0}
{x^{-r}y^{-s}\over{rs}}\bar c_r\cdot\bar c_s
\sum_{n,m\geq 0}(-z)^{n+m}
\left({-{1\over 3}\atop n}\right)
\left({-{2\over 3}\atop m}\right)
\bar c_{-{1\over 3}-n}\cdot\bar c_{
-{2\over 3}-m}|\widetilde
0\rangle\nonumber\\
&=&\sum_{r,s>0}x^{-r}y^{-s}\left({-{1\over 3}\atop{r-{1\over 3}}}\right)
\left({-{2\over 3}\atop{s-{2\over 3}}}\right)(-z)^{r+s-1}d\,,\end{eqnarray}
{\em i.e.}
\begin{equation}A(z,y,z)=-\sum_{r,s>0}\left({z\over x}\right)^r\left({z\over
y}\right)^s
\left({-{1\over 3}\atop{r-{1\over 3}}}\right)
\left({-{2\over 3}\atop{s-{2\over 3}}}\right)
d{(-z)^{r+s}\over{r+s}}\,.
\label{k6}
\end{equation}
(The constant of integration vanishes as $A(x,y,0)=0$.)
Therefore we have, from \reg{k3} and \reg{k6},
\begin{equation}A_{rs}=-\left({-{1\over 3}\atop{r-{1\over 3}}}\right)
\left({-{2\over 3}\atop{s-{2\over 3}}}\right)
{(-1)^{r+s}\over{r+s}}\,.
\label{appendix}
\end{equation}
Similarly, define
\begin{equation}B(x,y,z)=\sum_{r,r'>0}\langle\widetilde 0|{x^{-r}\over r}\bar
c_r^i
e^{zL_{-1}}{y^{r'}\over{r'}}\bar c^i_{-r'}|\widetilde
0\rangle\,,\end{equation}
(with summation over $i=1,\ldots,d$ implied).
Then from \reg{k1} we have
\begin{equation}
\label{k9}
B(x,y,z)=\sum_{r>0}{1\over r}\left({y\over x}\right)^rd+
\sum_{r,r'>0}\langle\widetilde 0|{x^{-r}\over r}\bar c_r^i
B_{rr'}\bar c_{-r}\cdot\bar c_{r'}{y^{r'}\over{r'}}\bar c^i_{-r'}|\widetilde
0\rangle\,,\end{equation}
(the higher order terms in the expansion of the exponentials in
\reg{k1} giving no
contribution as clearly at least two annihilation operators annihilate
$c^i_{-r'}|\widetilde 0\rangle$, and similarly $\langle\widetilde 0|c^i_r$
annihilates at least two creation operators), {\em i.e.}
\begin{equation}
\label{k10}
B(x,y,z)=\sum_{r>0}{1\over r}\left({y\over x}\right)^rd+
d\sum_{r,r'>0}B_{rr'}\left({z\over x}\right)^r
\left({y\over z}\right)^{r'}\,.\end{equation}
But
\begin{eqnarray}
{d\over{dz}}B(x,y,z)&=&\sum_{r,r'>0}
\langle\widetilde 0|{x^{-r}\over r}\bar c_r^i
e^{zL_{-1}}L_{-1}{y^{r'}\over{r'}}\bar c^i_{-r'}|\widetilde
0\rangle\nonumber\\
&=&\sum_{r,r'>0}
\langle\widetilde 0|{x^{-r}\over r}\bar c_r^i
e^{zL_{-1}}y^{r'}\bar c^i_{-r'-1}|\widetilde
0\rangle\,,\end{eqnarray}
the contribution from $L_{-1}|\widetilde 0\rangle$ vanishing by a similar
argument again.
Thus
\begin{eqnarray}
{d\over{dz}}B(x,y,z)&=&\sum_{r,r'>0\atop{n\geq 0}}
\langle\widetilde 0|{x^{-r}\over r}\bar c_r^i
y^{r'}
\left({-r'-1\atop n}\right)(-z)^n\bar c_{-r'-n-1}^i|\widetilde
0\rangle\nonumber\\
&=&\sum_{r'>0}\left({y\over x}\right)^{r'}
\left(1-{z\over x}\right)^{-r'-1}{1\over x}d\,.
\end{eqnarray}
So
\begin{equation}
\label{k13}
B(x,y,z)=\sum_{r'>0}\left({y\over x}\right)^{r'}
\left(1-{z\over x}\right)^{-r'}{1\over{r'}}d\,,\end{equation}
since $B(x,y,0)=\sum_{r>0}{1\over r}\left({y\over x}\right)^rd$ from \reg{k10}.
Expanding \reg{k13} gives
\begin{eqnarray}
B(x,y,z)&=&\sum_{r'>0\atop{r\geq r'}}\left({y\over x}\right)^{r'}
\left(-{z\over
x}\right)^{r-r'}\left({-r'\atop{r-r'}}\right){1\over{r'}}d\nonumber\\
&=&\sum_{r'>0\atop{r\geq r'}}\left({y\over z}\right)^{r'}
\left({z\over x}\right)^r(-1)^{r-r'}\left({-r'\atop{r-r'}}\right){1\over{r'}}d
\,,
\end{eqnarray}
and comparing with \reg{k9} we see that
\begin{equation}B_{rr'}=\cases{(-1)^{r-r'}\left({-r'\atop{r-r'}}\right){1\over{r'}} &
for $r>r'$\cr
0 & otherwise}\,.\end{equation}
Similarly
\begin{equation}C_{ss'}=\cases{(-1)^{s-s'}\left({-s'\atop{s-s'}}\right){1\over{s'}} &
for $s>s'$\cr
0 & otherwise}\,.\end{equation}
\section{Symmetry Between the Vertex Operators $W_3$
and $\Wbar_3$}
\label{App2}
The locality relations \reg{nine} and \reg{ten} together with the obvious
isomorphism between $W_1(\rho,\zeta)\Wbar_1(\chi,z)$ and
$W_2(\overline\rho,\zeta)\Wbar_2(\overline\chi,z)$ given by interchange
of the oscillators $c$ and $\bar c$ suggests a corresponding symmetry
between the vertex operators $W_3(\chi,z)$ and $\Wbar_3(\overline\chi,z)$,
which would be expected naively in any case. In this appendix, we shall
verify this symmetry, making use of the consistency of the $\ze_3$-twisted
conformal field theory.

For a consistent conformal field theory, $\Wbar_3$ must satisfy a relation
analogous to that defining $W_3$, {\em i.e.} \reg{n798}. Hence, we need
only check the symmetry of $W_3(\chi_0,z)\chi_0\equiv z^{-\Delta_0}
F'_{\chi_0}(z)$ and $\Wbar_3(\overline\chi_0,z)\overline\chi_0
\equiv z^{-\Delta_0}
\overline F'_{\chi_0}(z)$. From \reg{n789} we have
\begin{equation}\Wbar_3(\overline\chi_0,z)=z^{-2\Delta_0}{W_3}^\dagger(\chi_0,
1/z^\ast)\,.\end{equation}
Let $\psi$ be a state in the untwisted sector (with $\theta=1$ strictly).
Then, by the conjugate of \reg{seven},
\begin{equation}V_1(\psi,\zeta)\Wbar_3(\overline\chi_0,z)=
z^{-2\Delta_0}{W_3}^\dagger(\chi_0,
1/z^\ast)V_2(\psi,\zeta)\,,\end{equation}
and hence
\begin{equation}
\label{l3}
\langle\chi_0|V_1(\psi,\zeta)|\overline F'_{\chi_0}(z)\rangle=
\langle F'_{\chi_0}(1/z^\ast)|V_2(\psi,\zeta)|\overline\chi_0\rangle
\,.\end{equation}
Let us consider only the bilinear part of $\overline F'_{\chi_0}(z)$,
{\em i.e} we take
\begin{equation}\Wbar_3(\overline\chi_0,z)\overline\chi_0=
z^{-\Delta_0}\exp\left\{\sum_{r,s>0}c_{-r}\cdot c_{-s}
z^{r+s}\overline D_{rs}\right\}\overline\chi(\chi_0)\,,\end{equation}
We know that $\Wbar_3$ must satisfy a relation
analogous to \reg{n7117}, and therefore to prove coincidence of the generating
functions for $D_{rs}$ and $\overline D_{rs}$ we need only verify
equality for the diagonal forms ({\em c.f.} the argument given
for the derivation of an expression for the bilinear part of
$F_{\chi_0}(z)$), {\em i.e.} we must show
\begin{equation}\overline g(x,x)=g(x,x)\,,\end{equation}
where
\begin{equation}\overline g(x,y)\equiv\sum_{r,s>0}\overline
D_{rs}x^ry^s\,
\end{equation}
by analogy with the expression for $g$ given following (\reg{n7124}).

Consider $\psi=\psi_L$ in \reg{l3}. This gives us
\begin{equation}\sum_{r,s>0}rsd\overline D_{rs}z^{r+s}\zeta^{-r-s-2}+
{d\over 9}{1\over{\zeta^2}}=
\sum_{r,s>0}rsd{D_{rs}}^\ast z^{-r-s}\zeta^{r+s-2}+
{d\over 9}{1\over{\zeta^2}}\,,\end{equation}
{\em i.e.}, noting that the coefficients $D_{rs}$ which we
evaluated in the paper are real,
\begin{equation}\left({z\over\zeta}\right)^2\overline
h\left({z\over\zeta}\right)=
\left({\zeta\over z}\right)^2h\left({\zeta\over z}\right)\,,\end{equation}
where
\begin{equation}\left.h(a)\equiv{d\over{da}}{d\over{db}}g(a,b)\right|_{a=b}\,,
\end{equation}
and similarly for $\overline h$.

However, we see from \reg{n7125} and \reg{n7138}
that $a^2h(a)$ is invariant
under $a\mapsto 1/a$,
and so $\overline h(a)=h(a)$ as required. A similar argument gives the
analogous result for the lattice dependent piece of $F_{\chi_0}(z)$.

\end{document}